\documentclass
[superscriptaddress,twocolumn,secnumarabic,amssymb,amsmath,nobibnotes,aps,prd,showpacs,nofootinbib]{revtex4}%
\usepackage{graphicx}
\usepackage{epsf}
\usepackage{bm}
\usepackage{amsmath}
\usepackage{amsfonts}
\usepackage{amssymb}
\usepackage{epstopdf}%
\providecommand{\U}[1]{\protect\rule{.1in}{.1in}}

\newcommand{\be}{\begin{equation}}
\newcommand{\ee}{\end{equation}}

\newcommand{\mincir}{\raise
-3.truept\hbox{\rlap{\hbox{$\sim$}}\raise4.truept\hbox{$<$}\ }}
\newcommand{\magcir}{\raise
-3.truept\hbox{\rlap{\hbox{$\sim$}}\raise4.truept\hbox{$>$}\ }}

\usepackage{color}
\begin{document}

\title{Challenging bulk viscous unified scenarios with cosmological observations}

\author{Weiqiang Yang}
\email{d11102004@163.com}
\affiliation{Department of Physics, Liaoning Normal University, Dalian, 116029, P. R. China}

\author{Supriya Pan}
\email{supriya.maths@presiuniv.ac.in}
\affiliation{Department of Mathematics, Presidency University, 86/1 College Street, Kolkata 700073, India}

\author{Eleonora Di Valentino}
\email{eleonora.divalentino@manchester.ac.uk}
\affiliation{Jodrell Bank Center for Astrophysics, School of Physics and Astronomy, University of Manchester, Oxford Road, Manchester, M13 9PL, United Kingdom}

\author{Andronikos Paliathanasis}
\email{anpaliat@phys.uoa.gr}
\affiliation{Institute of Systems Science, Durban University of Technology, PO Box 1334,
Durban 4000, Republic of South Africa}

\author{Jianbo Lu}
\email{lvjianbo819@163.com}
\affiliation{Department of Physics, Liaoning Normal University, Dalian, 116029, P. R. China}

\begin{abstract}
In a spatially flat Friedmann-Lema\^{i}tre-Robertson-Walker universe, we investigate a unified cosmic fluid scenario endowed with bulk viscosity in which the coefficient of the bulk viscosity has a power law evolution. The power law  in the bulk viscous coefficient is a general case in this study which naturally includes several choices as  special cases. Considering such a general bulk viscous scenario, in the present work we have extracted the observational constraints using the latest cosmological datasets and examine their behaviour at the level of both background and perturbations. From the observational analyses, we find that a non-zero bulk viscous coefficient is always favored and some of the models in this series are able to weaken the current tension on $H_0$ for some dataset. However, from the Bayesian evidence analysis, $\Lambda$CDM is favored over the bulk viscous model.

\end{abstract}

\maketitle

\section{Introduction}

Observational evidences from a series of distinct astronomical sources firmly state that nearly 96\% of the total energy budget of the universe is comprised of the dark sector \cite{Ade:2015xua}. This dark sector is usually classified into dark matter and the dark energy where the dark matter is responsible for the structure formation of the universe and dark energy is speeding up  the expansion of the universe into an accelerating manner. 
The simplest possibility to model such dark universe has been proposed in terms of the non-interacting $\Lambda$CDM cosmology in which $\Lambda >0$, the cosmological constant, plays the role of dark energy and the dark matter sector is comprised with CDM (cold dark matter). However, the problems related to the cosmological constant have motivated the construction of alternative cosmological models. This resulted in a class of cosmological models where either the dark fluids evolve separately (known as non-interacting cosmological models) or the dark fluids have a mutual interaction between them, also known as interacting dark energy models (see \cite{Copeland:2006wr, Bolotin:2013jpa} for reviews on a class of non-interacting and interacting dark energy models). However, in spite of many investigations,  the origin, nature and the evolution of these dark fluids are absolutely unknown until today.  

Along the same line of investigations, a possible and natural idea in such a context is that the dark fluids, namely, the dark matter and dark energy, are not two exotic matter components rather they are just two different aspects
of a single fluid model, usually known as the {\it unified dark matter} (UDM) scenarios. Theoretically, there is  no objection to consider such UDM scenarios since the nature of dark sector could be anything. 
In the context of Einstein's gravitational theory, such UDM
scenarios are described by an equation of state $ p = f (\rho)$, where $p$ and $\rho$, are respectively the pressure and energy density of the UDM fluid and 
$f$ is any analytic function of the energy density, $\rho$ (one can quickly recall the Chaplygin gas \cite{Chaplygin,Kamenshchik:2001cp, Bilic:2001cg,Gorini:2002kf,gorini01,Bento:2002ps,Lu:2009zzf,Xu:2012qx,Benaoum:2002zs,Debnath:2004cd,Lu:2008zzb,Xu:2012ca} and other unified cosmologies \cite{Hova:2010na,Hernandez-Almada:2018osh,Yang:2019jwn} in this context). Sometimes, these UDM scenarios are also studied in the form of $p = g (H)$ where  $g$ is any analytic function of  
$H$, the Hubble rate of the Friedmann-Lema\^{i}tre-Robertson-Walker (FLRW) universe which is the well known geometrical description of our universe in the large scale and in this present work we have considered such geometrical configuration. However, one can see that the prescriptions, $p = f(\rho)$ and $p = g(H)$ are actually equivalent for a spatially flat universe since $\rho \propto H^2$ for this universe, however, for nonflat cases, they are not same. Following this one could make a number of choices for such UDM models keeping only one thing that both $f$ and $g$ should be analytic with respect to their corresponding arguments.   
A class of cosmological scenarios  with this equation of state has been investigated in detail in the context
of cosmological bulk viscosity \cite{Barrow:1990vx,Barrow:1988yc,Barrow:1986yf,Pavon:1990qf,Zimdahl:1996ka,Coley:1995uh,Gavrilov:1997mj,Chimento:1997ga,Belinchon:1998ij,Belinchon.:2001pj,Mak:2001wm,Brevik:2001ed,Mak:2001uj,Mak:2003jq,Brevik:2005bj,Giovannini:2005ii,Giovannini:2005af,Cataldo:2005qh,Fabris:2005ts,Hu:2005fu,Colistete:2007xi,Avelino:2008ph,Li:2009mf,HipolitoRicaldi:2009je,Velten:2011bg,Gagnon:2011id,Brevik:2011mm,Pourhassan:2013sw,Avelino:2013zxh,Avelino:2013wea,Velten:2013qna,Disconzi:2014oda,Mostaghel:2016lcd,Barbosa:2017ojt,Haro:2015ljc,Sasidharan:2018bay} (also see \cite{Brevik:2017msy} for a recent review on bulk viscous cosmologies).  A bulk viscous fluid is a cosmic fluid endowed with bulk viscosity. Effectively, a bulk viscous fluid with $(p, \rho)$ as respectively the pressure and energy density is identified with an effective pressure $p_{\rm eff} =  (\gamma -1 )\rho - \eta (\rho) u^{\mu}_{; \mu}$, in presence of the bulk viscosity \footnote{In this connection we recall an equivalent cosmological theory known gravitationally induced particle creation theory \cite{Chakraborty:2014ora,Chakraborty:2014fia,Pan:2014lua,Nunes:2015rea, Lima:2015xpa, deHaro:2015hdp, Pan:2016jli, Nunes:2016aup,Nunes:2016eab,Nunes:2016tsf, Paliathanasis:2016dhu, Salo:2016xlw,Pan:2016bug, Pan:2018ibu} which is equivalent to the bulk viscous theory at the level of equations but from the thermodynamical point of view both are theories are distinct \cite{Lima:1992np}.}. Here, $u^{\mu}_{; \mu}$ is the expansion scalar of this fluid, $\eta (\rho) >0$ is the coefficient of the bulk viscosity 
and $\gamma$ is the model parameter. One can identify $\gamma$ as 
a conventional equation of state of the fluid in absence of the bulk viscosity. For FLRW universe, $u^{\mu}_{; \mu} = 3H$, hence the effective  pressure of the bulk viscous fluid in this universe turns out to be  $p_{\rm eff} =  (\gamma -1 )\rho - 3 H \eta (\rho)$. In Refs. \cite{Barrow:1990vx,Barrow:1988yc,Barrow:1986yf,%
Pavon:1990qf,Zimdahl:1996ka,Fabris:2005ts,Colistete:2007xi, Li:2009mf} the viscous coefficient was taken to be $\eta (\rho) = \alpha \rho^m$, where $\alpha, m$ are free parameters and cosmic expansion was investigated in detail, but all the above works were mostly theoretical both at background and perturbations. Concerning the observational examinations,  although the low redshifts data like Supernovae Type Ia were encountered but the full cosmic microwave background (CMB) temperature and anisotropy data (we acknowledge that CMB shift parameter was introduced in \cite{Li:2009mf}), so far we are aware of the literature, have not been applied to such models. Thus, we believe that such an analysis will be worth for a complete picture of such scenarios.

Thus, in order to take into account the observational features of bulk viscous models, in the present work we have considered two specific UDM scenarios and constrained them using different cosmological data. We have also studied the evolution of these models at the level of perturbations through the temperature anisotropy in the CMB spectra and matter power spectra as well.

The work has been organized in the following manner. In section \ref{sec-efe} we 
present the gravitational field equations for an imperfect fluid with bulk viscosity. 
In section \ref{sec-data} we present the observational data and the constraints of the present models. In particular, in the subsections \ref{sec-bvf1} and \ref{sec-bvf2} we respectively summarize the main results of the two models. Finally, we close the present work in section \ref{sec-conclu} with a brief summary.

\section{The background and perturbation equations for a viscous universe}
\label{sec-efe}

We consider a homogeneous and isotropic model of our universe which is characterized 
by the usual Friedmann-Lema\^{i}tre-Robertson-Walker line element $$ds^2 = -dt^2 + a^2 (t) \left[\frac{dr^2}{1-kr^2} + r^2 \left(d \theta^2 + \sin^2 \theta d \phi^2\right)\right],$$
where $a(t)$ is the scale factor of the universe; $k$ is the spatial curvature which for its three distinct values, namely, $0$, $-1$, $+1$, respectively represent a spatially flat, open and a closed universe.  
In this work we shall confine ourselves to the spatially flat universe, that means, $k =0$ throughout the work. The energy density, $\rho$, of the universe in this spacetime is thus constrained by the Hubble rate $H \equiv \dot{a}/a$ as ($8 \pi G =1 $)

\begin{eqnarray}\label{hubble}
3 H^2 =  \rho_t = (\rho_r + \rho_b + \rho_D) \,,
\end{eqnarray}
where $\rho_r$, $\rho_b$, $\rho_D$ are respectively the energy density for the radiation and baryons and the unified fluid where  $p_D = (\gamma -1) \rho_D$ and we have a bulk viscosity background. 
Following  \cite{Barrow:1988yc}, the effective pressure of the unified dark fluid can be written as 

\begin{eqnarray}\label{eff-pressure}
p_{\rm eff} = p_D - 3 H \eta (\rho_D)
\end{eqnarray}
where $\eta (\rho_D)$ is the coefficient of the bulk viscosity that we assume to take the  following well known form \cite{Barrow:1990vx,Barrow:1988yc,Barrow:1986yf}: 

\begin{eqnarray}\label{BVF}
\eta (\rho_D)=\alpha\rho_D^m,\;\;\; (\alpha>0). 
\end{eqnarray}
Thus, the effective pressure for the unified dark fluid can be recast into 
\begin{eqnarray}\label{eq2}
p_{\rm eff}=(\gamma-1)\rho_D -3\alpha H\rho_D ^m,
\end{eqnarray}
In the spatially flat universe, at late time the contributions from radiation and baryons are negligible, so, approximately, $H\propto\rho_D^{1/2}$, this is equivalent to a total stress of the form
$p_{\rm eff} \approx (\gamma-1)\rho_D-\sqrt{3}\alpha\rho_D^{m+1/2}$ \cite{Barrow:1990vx,Barrow:1988yc,Barrow:1986yf},
where this equation reduces to the generic form $p_{\rm eff}+\rho_D \approx \Gamma \rho_D^{\delta}$ with $\gamma= 0$ and $\delta=m+1/2$ and $\Gamma = -\sqrt{3}\alpha$. For these kind of effective pressure of the dark fluid, there are three free model parameters $\gamma$, $\alpha$, $m$. The pressure $p_{\rm eff}=-3\alpha H\rho_D^m$ with assuming $\gamma=1$, and we can define the dimensionless parameter $\beta=\alpha H_0\rho_{t0}^{m-1}$, where $\rho_{t0}$ is the present value of $\rho_t$ defined in eqn.~(\ref{hubble}). Let us note that 
the model with $p_{eff}=-3\alpha H\rho_D^m$ was constrained by the Supernovae Type Ia \cite{Riess:2004nr} and CMB shift-parameter \cite{Wang:2006ts} data where the best-fit values of the parameters were found to be $m=-0.4$ and $\beta=0.236$.

In the present work we seek for a robust observational analysis of the bulk viscous cosmologies  \cite{Barrow:1990vx,Barrow:1988yc,Barrow:1986yf,%
Pavon:1990qf,Zimdahl:1996ka,Li:2009mf} in a spatially flat universe. Thus, in presence of the bulk viscosity, the effective pressure of the unified fluid becomes,
\begin{eqnarray}\label{eq3}
p_{\rm eff}=(\gamma-1)\rho_D-\sqrt{3}\alpha\rho_t^{1/2}\rho_D^m.
\end{eqnarray}
where the baryons and radiation components present in $\rho_t$ (see eqn.~\ref{hubble}) are conserved separately, and hence they evolve as, 
$\rho_b=\rho_{b0}a^{-3}$, $\rho_r=\rho_{r0}a^{-4}$, respectively.
Here, the energy density of viscous dark fluid would not be in an analytical form due to the baryons and radiation in the effective pressure, however it could be solved numerically.
The effective equation of state of viscous dark fluid will be
\begin{eqnarray}\label{eq3}
w_{\rm eff}=(\gamma-1)-\sqrt{3}\alpha\rho_t^{1/2}\rho_D^{m-1}.
\end{eqnarray}

The adiabatic sound speed for the viscous fluid is
\begin{equation}
c_{a,\rm eff}^{2}=\frac{p_{\rm eff}^{\prime}}{\rho^{\prime}}=w_{\rm eff}+\frac{w_{\rm eff}^{\prime}}{3\mathcal{H}(1+w_{\rm eff})}.
\end{equation}
where the prime denotes the derivative of the conformal time. $\mathcal{H}$ is the conformal Hubble parameter, $\mathcal{H}=aH$.

When the equation of state of a purely barotropic fluid is negative, it has an imaginary adiabatic sound speed which possibly causes instability of the perturbations. In order to avoid this problem, we will allow an entropy perturbation (non-adiabatic perturbation) in the dark fluid according to the analysis of generalized dark matter \cite{Hu:1998kj}.

To follow the analysis of entropy perturbation for a generalized dark matter \cite{Hu:1998kj}, in the entropy perturbation mode, the true pressure perturbation is from the effective pressure, $p_{\rm eff}=p_D-3H\eta (\rho_D)$ is from $p_{\rm eff}=p_D-\eta(\nabla_\sigma u^\sigma)$, so one could calculate the pressure perturbation
\begin{eqnarray}
\delta p_{\rm eff}&=&\delta p_D-\delta\eta(\nabla_{\sigma}u^{\sigma})-\eta(\delta\nabla_{\sigma}u^{\sigma})\notag\\
&=&\delta
p_D-3H\delta\eta-\frac{\eta}{a}\left(\theta+\frac{h'}{2}\right).
\end{eqnarray}

Combined with $\eta=\alpha\rho_D^m$ \cite{Barrow:1986yf,Barrow:1988yc}, the effective sound speed of viscous dark fluid could be defined as
\begin{eqnarray}
&&c_{s,\rm eff}^{2}\equiv\frac{\delta p_{\rm eff}}{\delta \rho_D}|_{rf} \nonumber \\
&&=c_{s}^{2}-\sqrt{3}\alpha m\rho_t^{1/2}\rho_D^{m-1}-\frac{\alpha\rho_D^{m-1}}{a \delta_D}\left(\theta+\frac{h^{\prime}}{2}\right),
\end{eqnarray}
where '$|_{rf}$' denotes the rest frame, generally the sound speed $c^2_s=0$ in the rest frame according to the analysis of \cite{Hu:1998kj}.

To follow the formalism for a generalized dark matter \cite{Hu:1998kj}, one can write the perturbation equations of density contrast and velocity divergence
\begin{eqnarray}
\delta_D^{\prime}&=&-(1+w_{\rm eff})(\theta_D+\frac{h^{\prime}}{2})+\frac{w_{\rm eff}^{\prime}}{1+w_{\rm eff}}\delta_D \notag\\
&-&3\mathcal{H}(c_{s,\rm eff}^{2}-c_{a,eff}^{2})\left[\delta_D+3\mathcal{H}(1+w_{\rm eff})\frac{\theta_D}{k^{2}}\right]
,\\\
\theta_D^{\prime}&=&-\mathcal{H}(1-3c_{s,\rm eff}^{2})\theta_D+\frac{c_{s,\rm eff}^{2}}{1+w_{\rm eff}}k^{2}\delta_D,
\end{eqnarray}
\begin{figure}
\includegraphics[width=0.4\textwidth]{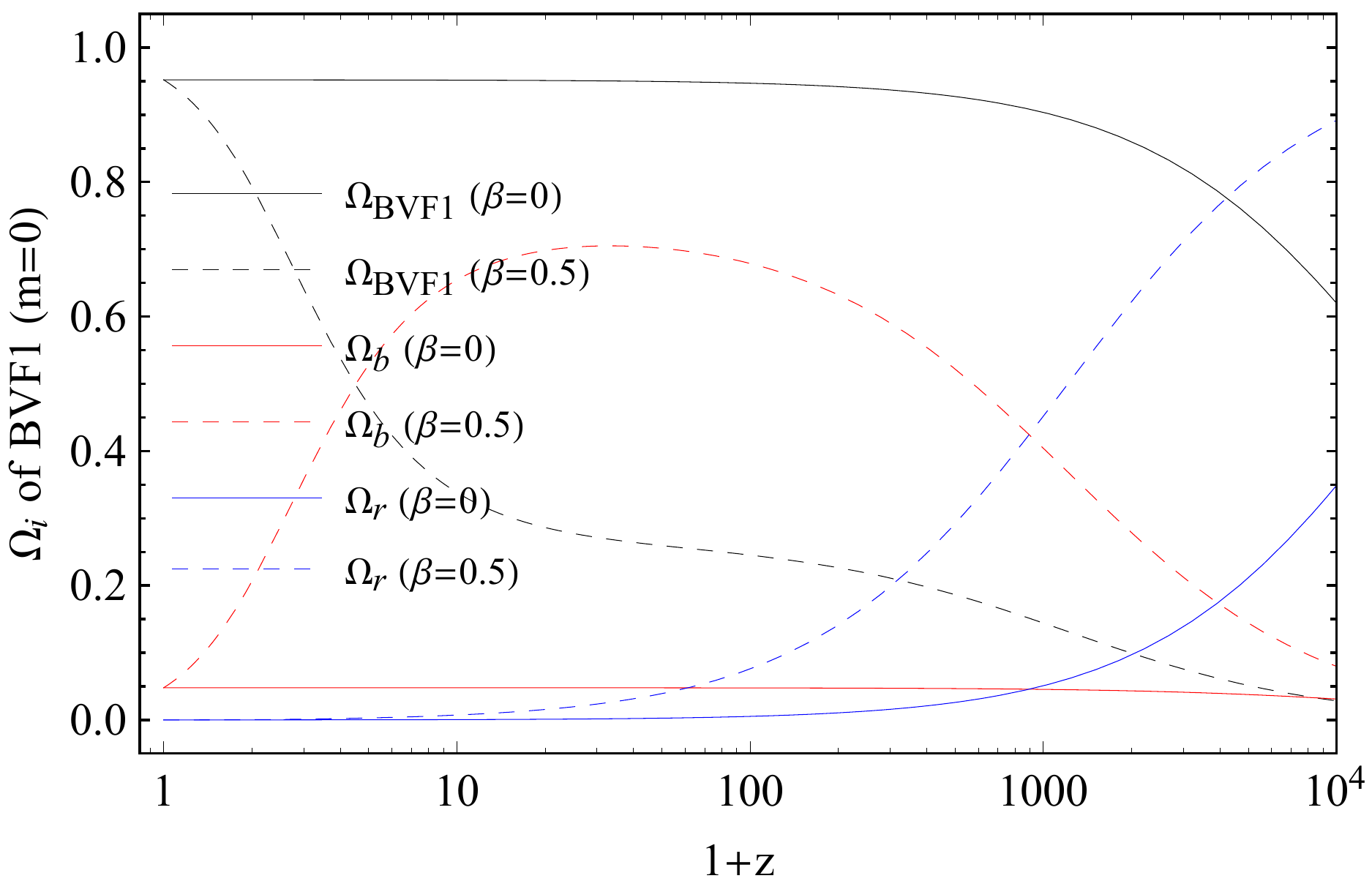}
\includegraphics[width=0.4\textwidth]{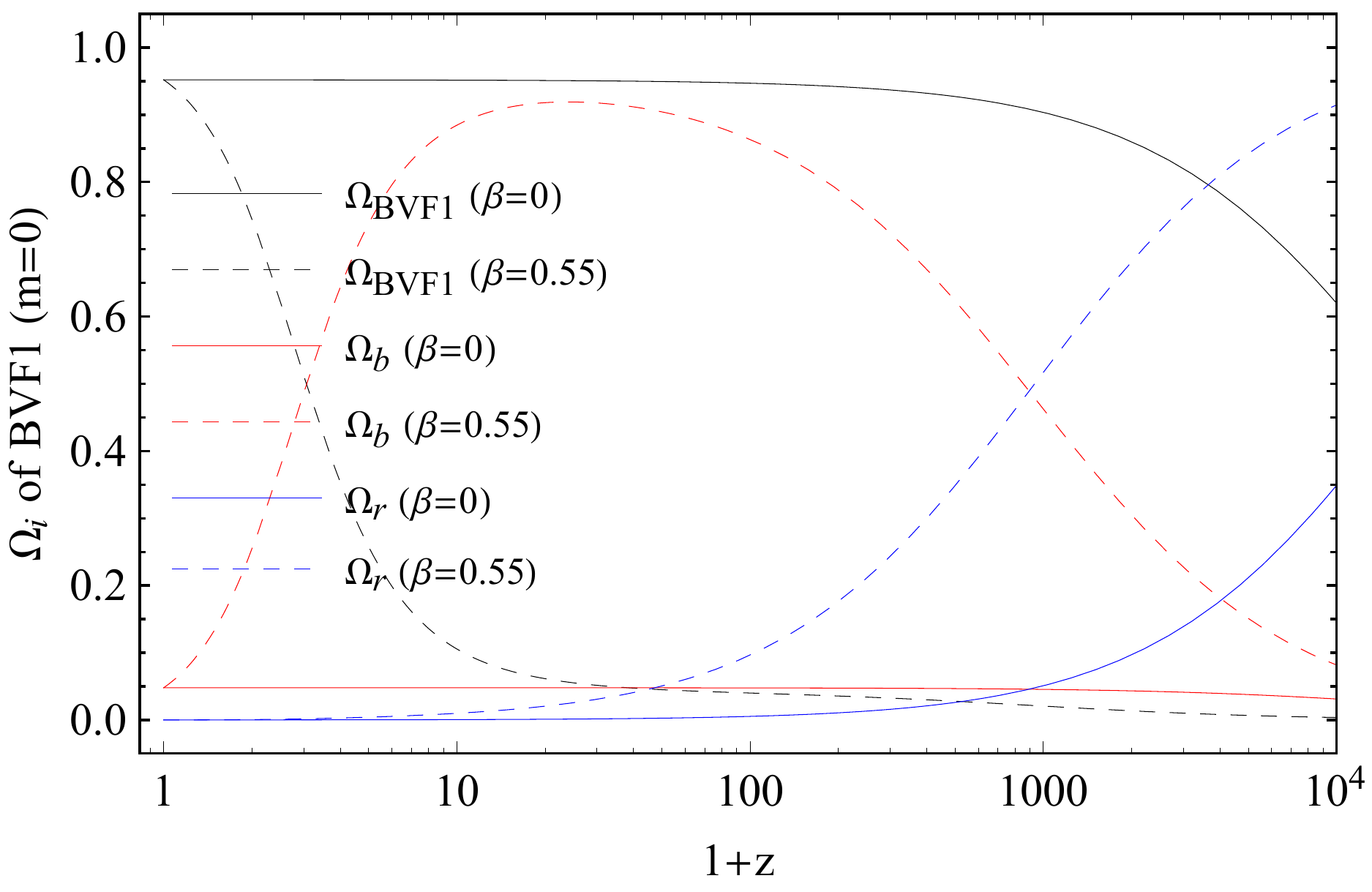}
\includegraphics[width=0.4\textwidth]{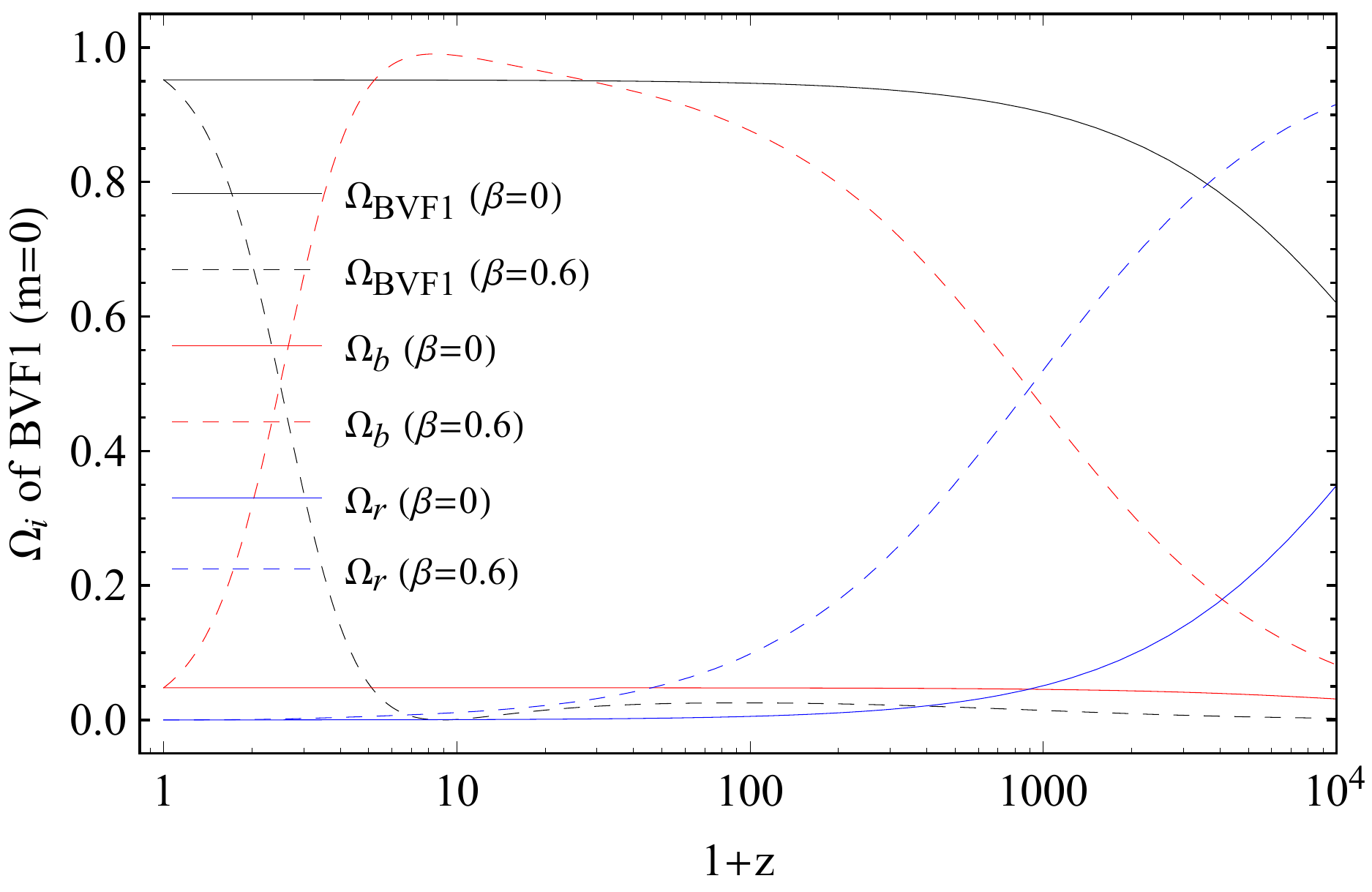}
\caption{Qualitative evolution of the density parameters for the BVF1 model with $m =0$ have been shown for different values of $\beta$, namely, $\beta =0.5$ (upper panel), $\beta =0.55$ (middle panel), $\beta = 0.6$ (lower panel), and also compared with no bulk viscous scenario (corresponding to $\beta =0$). }
\label{fig-Omega-bvf1-m0}
\end{figure}
\begin{figure}
\includegraphics[width=0.4\textwidth]{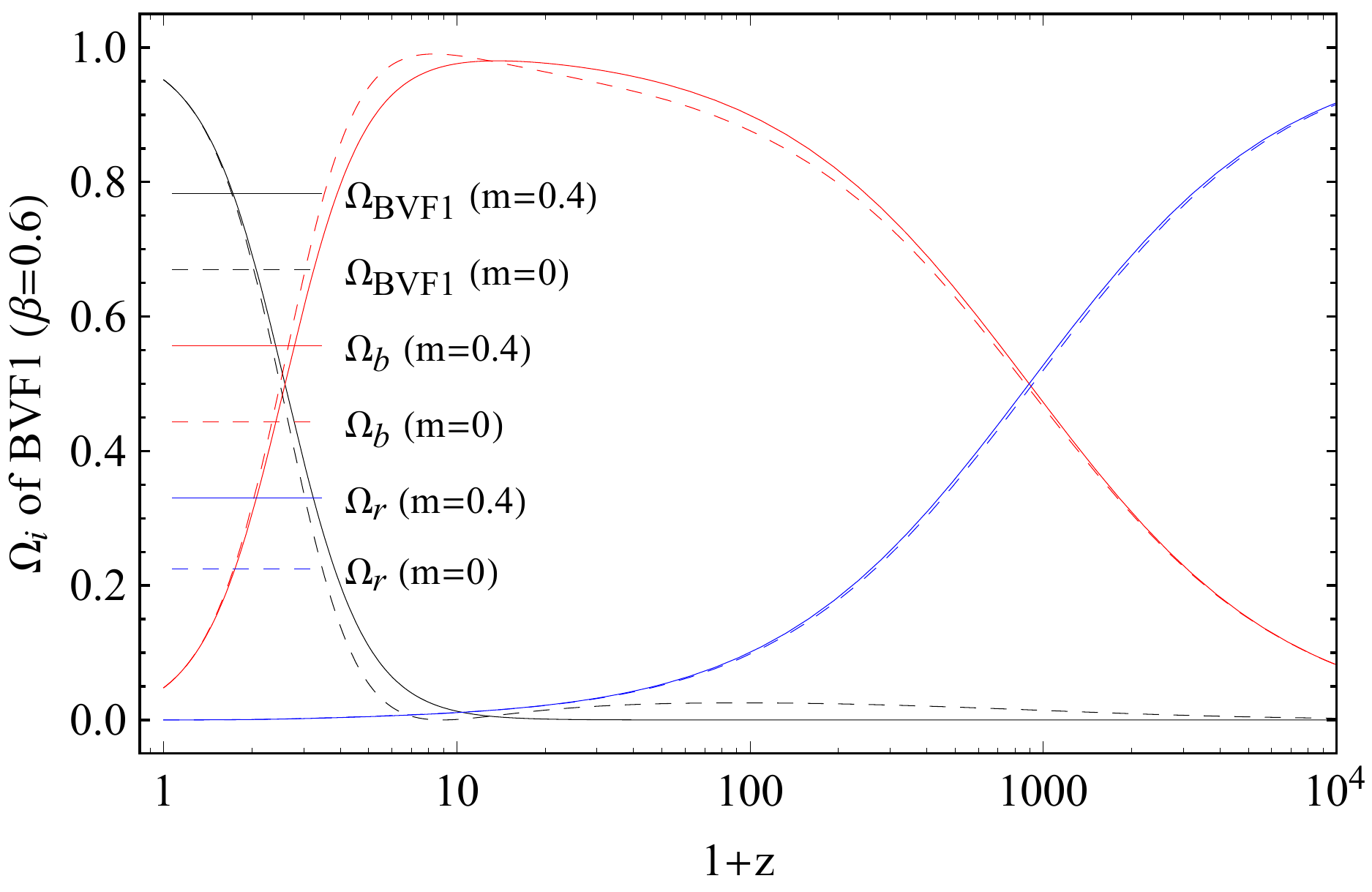}
\includegraphics[width=0.4\textwidth]{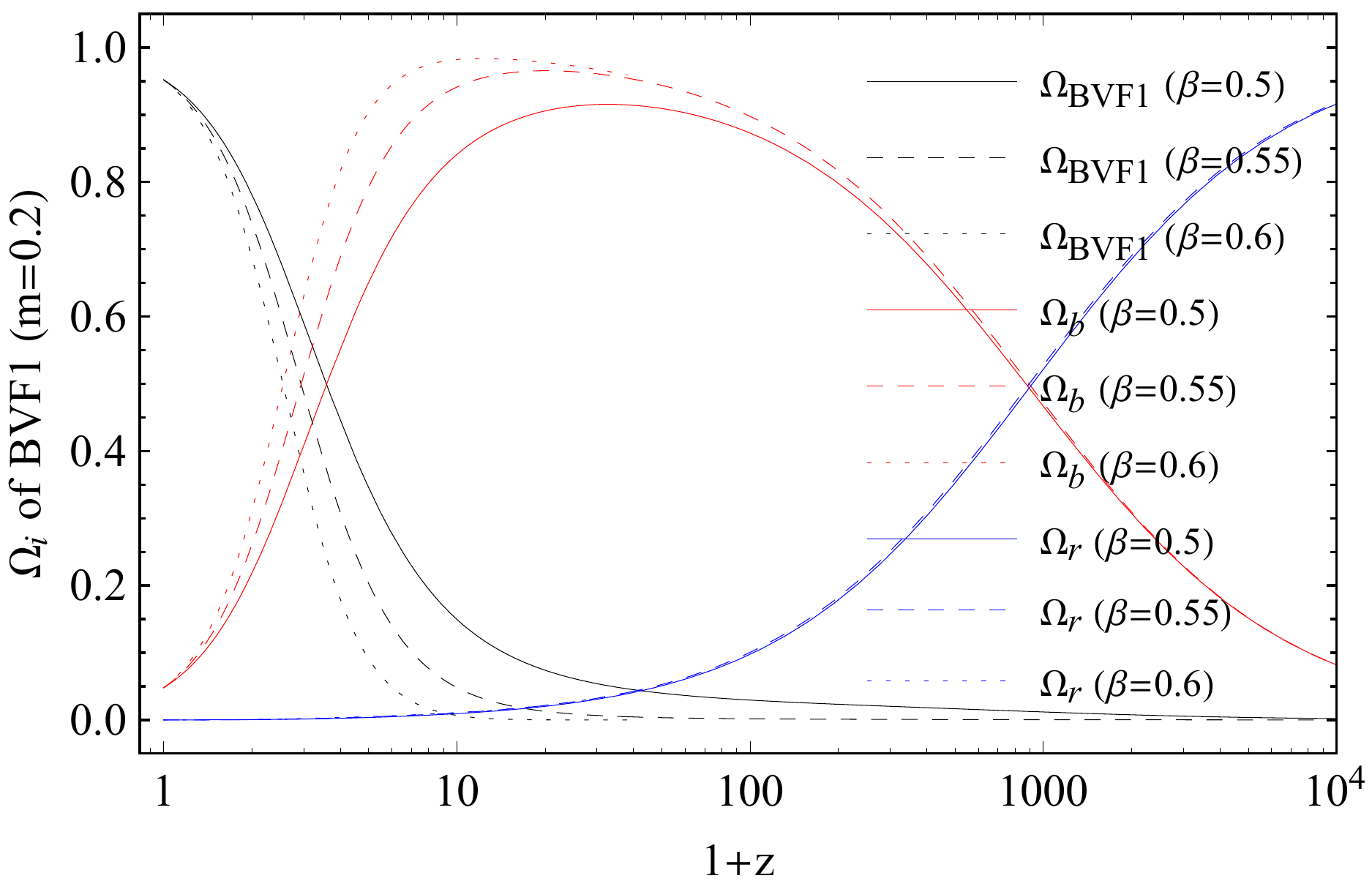}
\caption{We show some general behaviour of the BVF1 model considering the fact that $m \neq 0$. In the upper panel we fix $\beta  =0.6$ and consider the density parameters for $m =0.4$ and also compared to the constant bulk viscous scenario (corresponding to $m =0$). In the lower panel we fix $m  =0.2$ and consider three different values of $\beta$ in order to depict the evolution of the density parameters. }
\label{fig-Omega-bvf1-free-m}
\end{figure}
\begin{figure}
\includegraphics[width=0.4\textwidth]{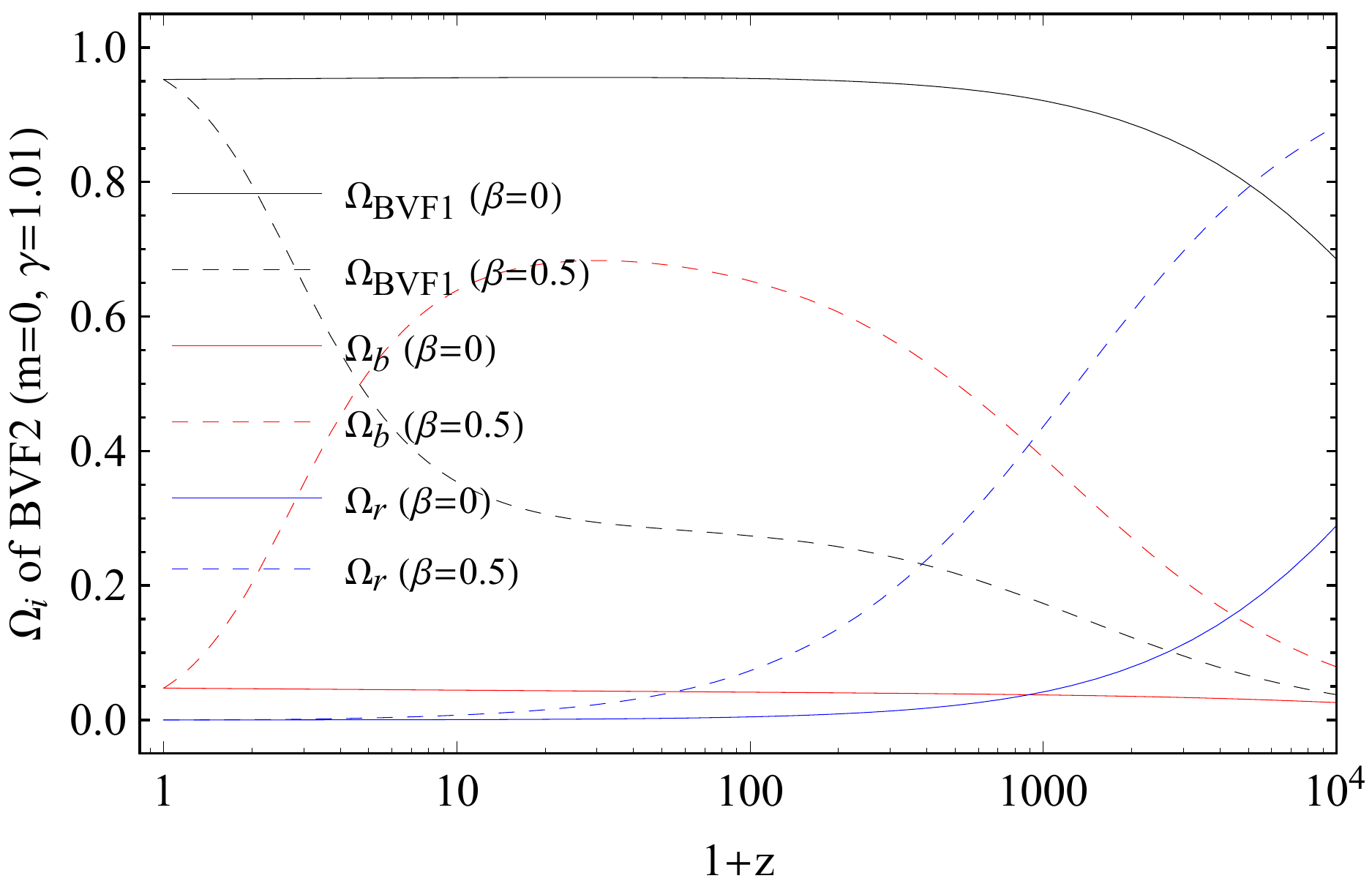}
\includegraphics[width=0.4\textwidth]{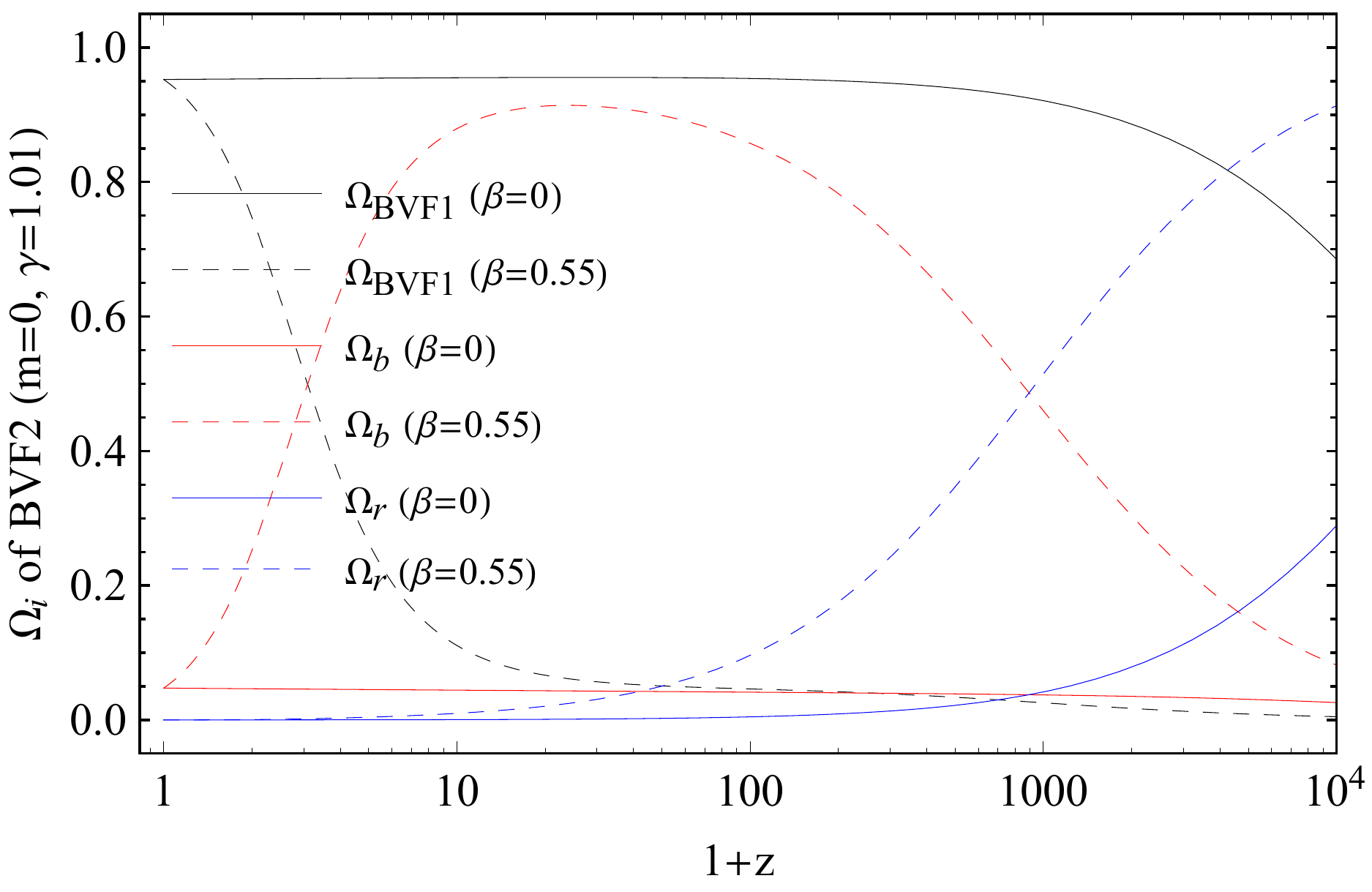}
\includegraphics[width=0.4\textwidth]{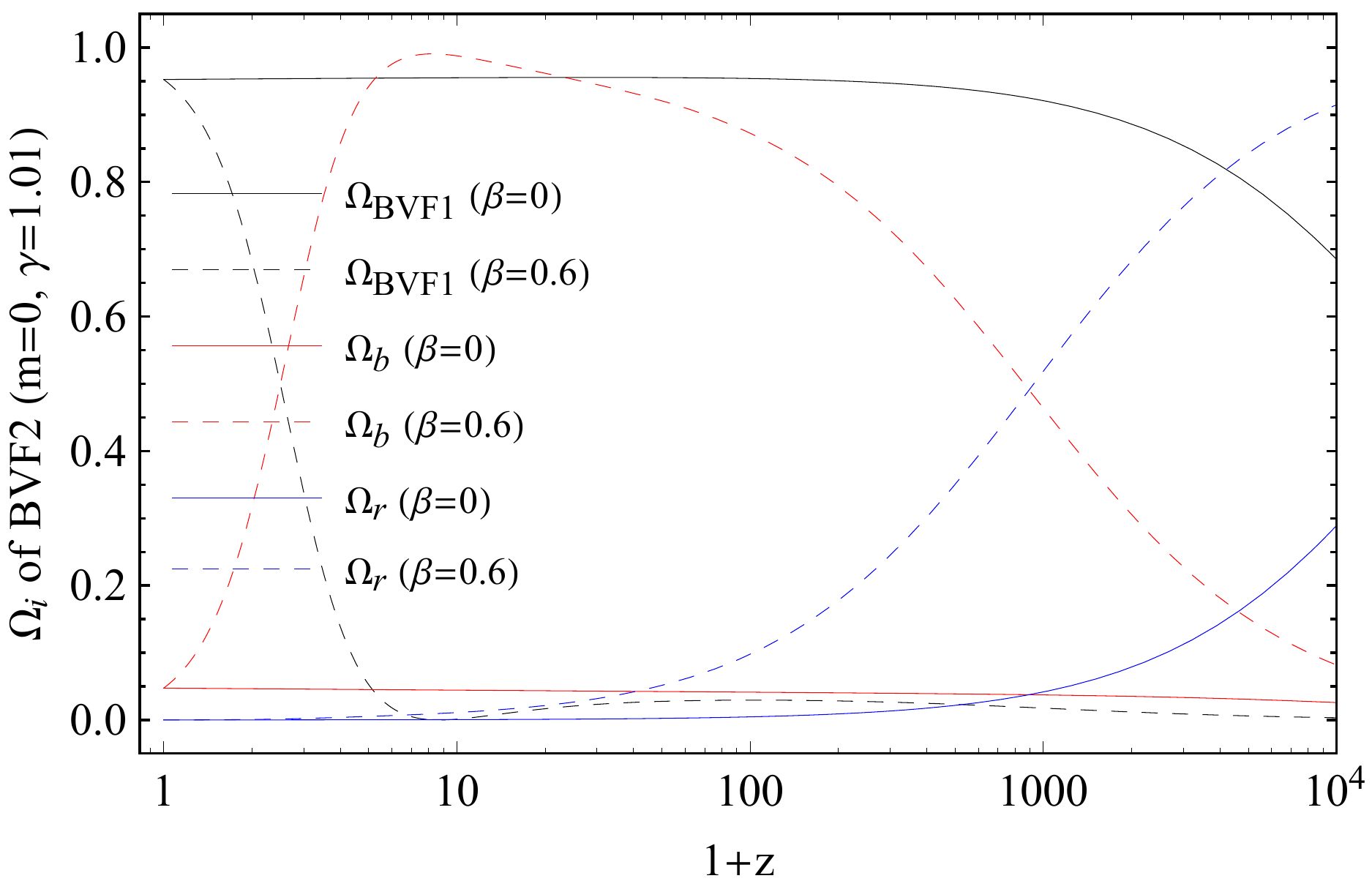}
\caption{Qualitative evolution of the density parameters for the BVF2 model with $m =0$ have been shown for different values of $\beta$, namely, $\beta =0.5$ (upper panel), $\beta =0.55$ (middle panel), $\beta = 0.6$ (lower panel), and also compared with no bulk viscous scenario (corresponding to $\beta =0$). Let us note that for all the plots we have fixed $\gamma = 1.01$. }
\label{fig-Omega-bvf2-m0}
\end{figure}
\begin{figure}
\includegraphics[width=0.4\textwidth]{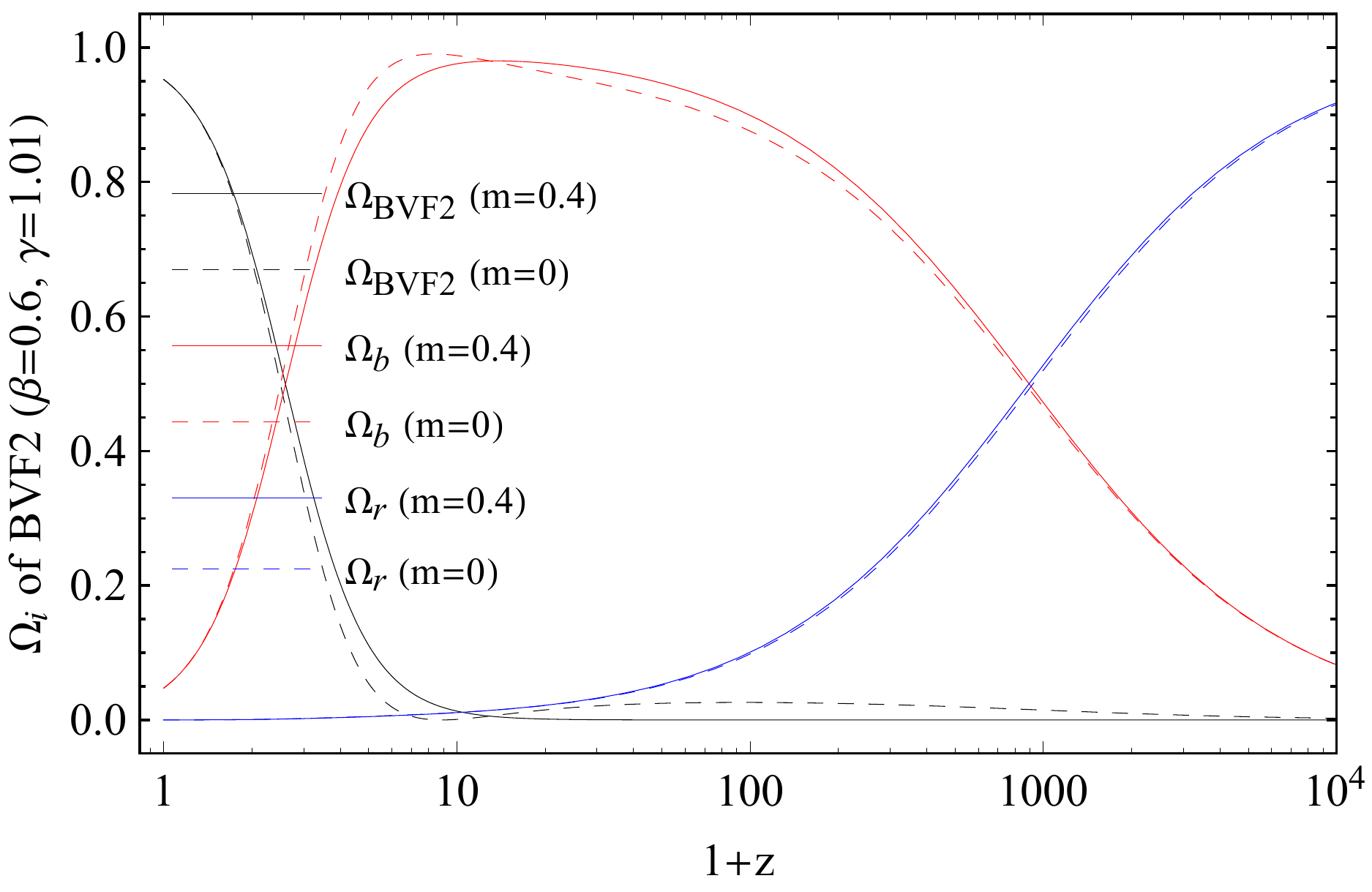}
\includegraphics[width=0.4\textwidth]{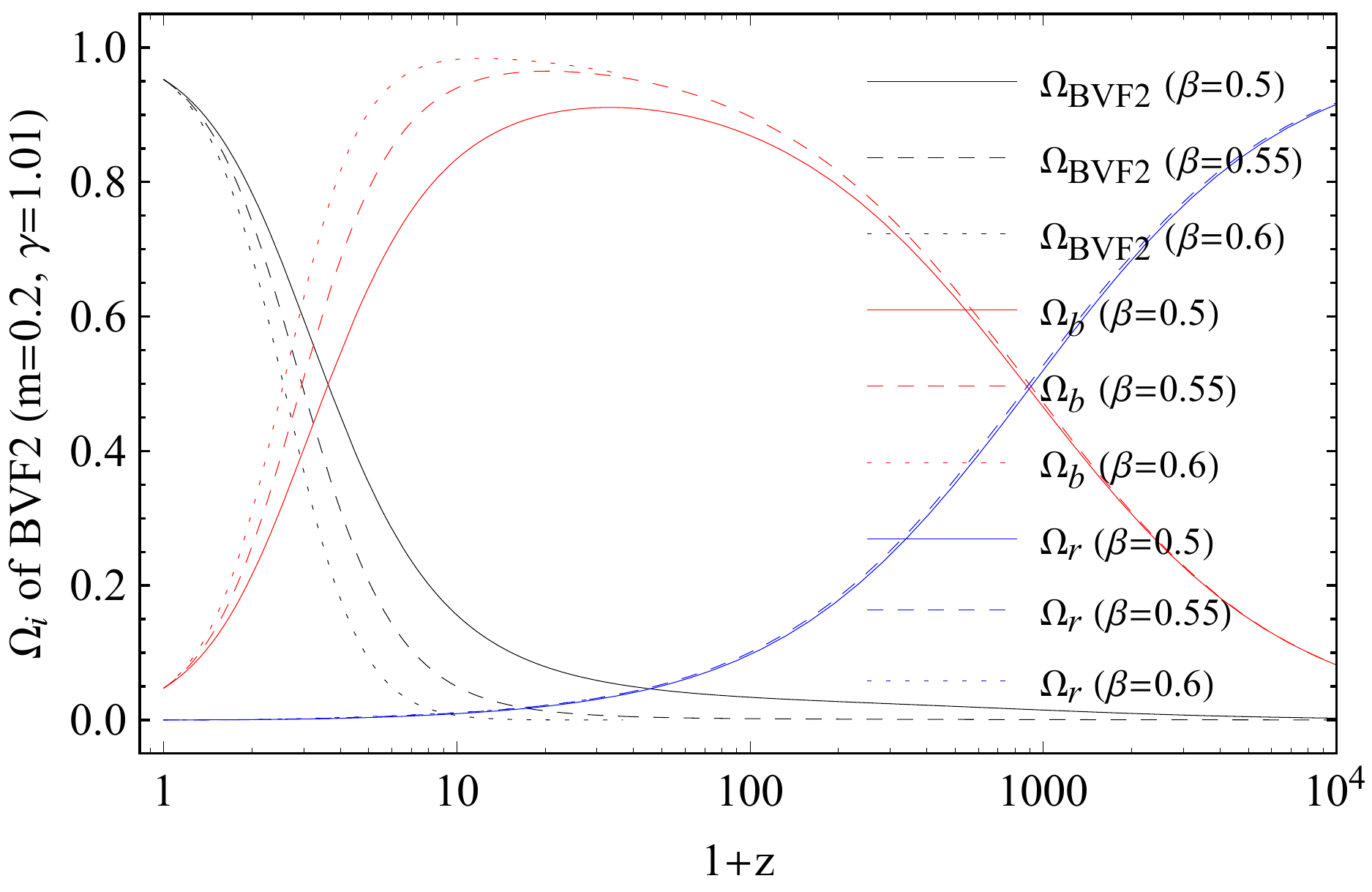}
\caption{We show some general behaviour of the BVF2 model considering the fact that $m \neq 0$. In the upper panel we fix $\beta  =0.6$ and consider the density parameters for $m =0.4$ and also compared to the constant bulk viscous scenario (corresponding to $m =0$). In the lower panel we fix $m  =0.2$ and consider three different values of $\beta$ in order to depict the evolution of the density parameters. Let us note that for all the plots we have fixed $\gamma = 1.01$.  }
\label{fig-Omega-bvf2-free-m}
\end{figure}

Now, based on the effective pressure of viscous dark fluid model \cite{Barrow:1990vx,Barrow:1988yc,Barrow:1986yf}, we consider two bulk viscous fluid models, namely the model with two free parameters $\alpha$ and $m$, labelled as BVF1. Another model with three parameters $\gamma$, $\alpha$, and $m$, labelled as BVF2. Let us note that for BVF1 and BVF2 models, for the purpose of statistical analysis, we have turned $\alpha$ into a 
a dimensionless quantity by defining $\beta=\alpha\rho_{0}^{m-1/2}$ in terms of the original parameter $\alpha$. Thus, from now on, we shall recognize $\beta$, $m$ as the governing parameters of Model BVF1 and the model BVF2 will be recognized by the 
parameters  $\beta$, $\gamma$ and $m$. The case with $m =0$ is the simplest bulk viscous scenario representing the constant bulk viscosity. Thus, in the present work we consider four different bulk viscous scenarios as follows: the two cases with $m =0$, that means, we consider two different scenarios, namely, BVF1 ($m=0$), and BVF2 ($m=0$), and secondly, we consider the general scenarios where $m$ acts as a free parameter, that means, the two cases for free $m$, named as BVF1 ($m$: free), and BVF2 ($m$: free). Now, in order to understand the qualitative evolution of the density parameters for radiation, baryons, and the effective bulk viscous fluid, we have systematically investigated all the possibilities. In Fig.  \ref{fig-Omega-bvf1-m0} we have shown the density parameters for the model BVF1 ($m =0$) using different values of $\beta$ such as $\beta = 0.5$ (upper panel of Fig.  \ref{fig-Omega-bvf1-m0}), $\beta =0.55$ (middle panel of Fig.  \ref{fig-Omega-bvf1-m0}) and $\beta  =0.6$ (lower panel of Fig.  \ref{fig-Omega-bvf1-m0}). From this figure (Fig.  \ref{fig-Omega-bvf1-m0}), we see that as $\beta$ increases,  the domination of the bulk viscous fluid starts lately.  Now, in order to understand the general scenario with free $m$, in Fig. \ref{fig-Omega-bvf1-free-m} we have depicted two different scenarios for the density parameters for some fixed values of $\beta$ and $m$. 
In a similar fashion, we have investigated the qualitative evolution of the density parameters for BVF2 considering both the possibilities, that means the case with $m =0$ (see Fig. \ref{fig-Omega-bvf2-m0}) and with free $m$ (see Fig. \ref{fig-Omega-bvf2-free-m}). From the qualitative evolution of various bulk viscous models presented in the aforementioned figures, one can easily notice that the dynamics associated with $\beta = 0$ is problematic since for $\beta  = 0$, one can see that (see Figs. \ref{fig-Omega-bvf1-m0} and \ref{fig-Omega-bvf2-m0}), at early time radiation was sub-dominated than the bulk viscous fluid, which is impossible.  
However, the above discussions imply that $\beta $ should be greater than zero in order to have realistic bulk viscous scenarios.

\section{Observational data and the results}
\label{sec-data}

In this section we describe both the observational data and the analyses of the present bulk viscous scenarios. In what follows, we first describe the observational datasets. 

\begin{itemize}

\item {\it Cosmic Microwave Background:} Cosmic Microwave Background (CMB) radiation is the effective astronomical probe to analyse the dark energy models. Here we consider the CMB temperature and polarization anisotropies together with their cross-correlations from Planck 2015 \cite{Adam:2015rua}. Particularly, we consider the  combinations of high- and low-$\ell$ TT likelihoods in the multipoles range $2\leq \ell \leq 2508$ and the combinations  of the high- and low-$\ell$ polarization likelihoods as well as \cite{Aghanim:2015xee}. 

\item {\it Pantheon sample from the Supernovae Type Ia data:} We use the most recent compilation of the supernovae type Ia (SNIa) comprising 1048 data points \cite{Scolnic:2017caz} in the redshift range $z \in [0.01, 2.3]$.

\item {\it Hubble parameter measurements:} Finally, we consider  the Hubble parameter values at different redshifts measured from the Cosmic Chronometers (CC). The Cosmic Chronometers are the most massive and passively evolving galaxies in the universe. For a  detailed motivation and measurements of the Hubble parameter values from CC, we refer to \cite{Moresco:2016mzx}. In this work we consider 30 measurements of the Hubble parameter values spread in the interval  $0< z < 2$, see again \cite{Moresco:2016mzx} where the data points are tabulated.

\end{itemize}

Now, in order to constrain the bulk viscous scenarios we have made use of the fastest algorithm, the Markov Chain Monte Carlo package \texttt{cosmomc} \cite{Lewis:2002ah} where an efficient convergence diagnostic, namely the Gelman-Rubin criteria  $R-1 $ exists that enables us to understand the convergence of the monte carlo chains.  For the first model BVF1, the analysed parameters space is, $\mathcal{P}_{\rm BVF1} =\{\Omega_b h^2, 100\theta_{MC}, \tau, n_s, {\rm{ln}}(10^{10} A_s), \beta, m\}$ and for the second model BVF2, the parameters space is, $\mathcal{P}_{\rm BVF2} =\{\Omega_b h^2, 100\theta_{MC}, \tau, n_s, {\rm{ln}}(10^{10} A_s), \beta, m, \gamma\}$ where $\Omega_bh^2$ is the baryons density, $100 \theta_{MC}$ is the ratio of the sound horizon to the angular diameter distance; $\tau$ is the
optical depth, $n_s$ is the scalar spectral index, $A_S$ is the amplitude of the initial power spectrum. 
In Table \ref{tab-priors}, we summarize the priors on the model parameters that have been used during the statistical analysis.  Let us now analyse the results of the models extracted from the observational datasets.

\begin{table}
\begin{center}
\begin{tabular}{c|c}
Parameter                    & Prior\\
\hline 
$\Omega_{b} h^2$             & $[0.005,0.1]$\\
$\Omega_{c} h^2$             & $[0.01,0.99]$\\
$\tau$                       & $[0.01,0.8]$\\
$n_s$                        & $[0.5, 1.5]$\\
$\log[10^{10}A_{s}]$         & $[2.4,4]$\\
$100\theta_{MC}$             & $[0.5,10]$\\ 
$\beta$                      & $[0, 1]$\\ 
$m$                          & $[-2, 0.5]$\\
$\gamma$                     & $[-3, 3]$  
\end{tabular}
\end{center}
\caption{We show the priors on the free parameters of the bulk viscous scenarios. }
\label{tab-priors}
\end{table}

\begingroup                                                                                                                     
\squeezetable                                                                                                                   
\begin{center}                                                                                                                  
\begin{table*}                                                                                                                   
\begin{tabular}{cccccccccccccc}                                                                                                            
\hline\hline                                                                                                                    
Parameters & CMB & CMB+CC & CMB+Pantheon & CMB+Pantheon+CC  \\ \hline

$\Omega_b h^2$ & $    0.02222_{-    0.00014-    0.00029}^{+    0.00015+    0.00029}$ & $    0.02227_{-    0.00017-    0.00031}^{+    0.00016+    0.00031}$ & $    0.02251_{-    0.00015-    0.00030}^{+    0.00016+    0.00030}$ &  $    0.02256_{- 0.00017-    0.00030}^{+    0.00015+    0.00031}$ \\

$100\theta_{MC}$ & $    1.03329_{-    0.00028-    0.00057}^{+    0.00028+    0.00055}$ & $    1.03332_{-    0.00027-    0.00053}^{+    0.00028+    0.00055}$ & $    1.03325_{-    0.00026-    0.00053}^{+    0.00027+    0.00052}$ & $    1.03327_{-    0.00033-    0.00057}^{+    0.00029+    0.00061}$ \\

$\tau$ & $    0.077_{-    0.016-    0.032}^{+    0.016+    0.033}$ & $    0.080_{-    0.017-    0.033}^{+    0.019+    0.031}$ & $    0.097_{-    0.018-    0.034}^{+    0.019+    0.035}$ & $    0.099_{-    0.017-    0.033}^{+    0.017+    0.032}$ \\

$n_s$ & $    0.9640_{-    0.0044-    0.0085}^{+    0.0043+    0.0088}$ & $    0.9652_{-    0.0045-    0.0083}^{+    0.0044+    0.0089}$ &  $    0.9741_{-    0.0043-    0.0080}^{+    0.0043+    0.0085}$ & $    0.9750_{-    0.0043-    0.0081}^{+    0.0043+    0.0087}$ \\

${\rm{ln}}(10^{10} A_s)$ & $    3.090_{-    0.032-    0.062}^{+    0.032+    0.062}$ & $    3.094_{-    0.031-    0.065}^{+    0.039+    0.059}$ & $    3.120_{-    0.035-    0.066}^{+    0.034+    0.069}$ & $    3.124_{-    0.033-    0.066}^{+    0.035+    0.062}$ \\

$\beta$ & $    0.199_{-    0.0042-    0.0084}^{+    0.0041+    0.0081}$ & $    0.201_{-    0.0044-    0.0092}^{+    0.0045+    0.0089}$ & $    0.211_{-    0.0048-    0.0080}^{+    0.0044+    0.0085}$ & $    0.212_{-    0.0040-    0.0075}^{+    0.0040+    0.0079}$ \\

$H_0$ & $   54.99_{-    0.30-    0.59}^{+    0.29+    0.59}$ & $   55.10_{-    0.31-    0.65}^{+    0.32+    0.63}$ & $   55.86_{-    0.35-    0.59}^{+    0.32+    0.63}$ & $   55.95_{-    0.33-    0.56}^{+    0.30+    0.60}$ \\
\hline 
$\chi^2_{\rm best-fit}$ & 12962.552 &  12991.452 & 14137.924 & 14163.750\\
\hline\hline                                                                                                                    
\end{tabular}                                                                                                                   
\caption{68\% and 95\% c.l. constraints on various free parameters of the model BVF1 assuming the simplest case $m =0$, that means, the constant bulk viscosity, using different observational data. Here $H_0$ is in the units of km/Mpc/sec.}
\label{tab:bvf1-m0}                                                                                                   
\end{table*}                                                                                                                     
\end{center}                                                                                                                    
\endgroup                                                                                                                       
\begin{figure*}
\includegraphics[width=0.7\textwidth]{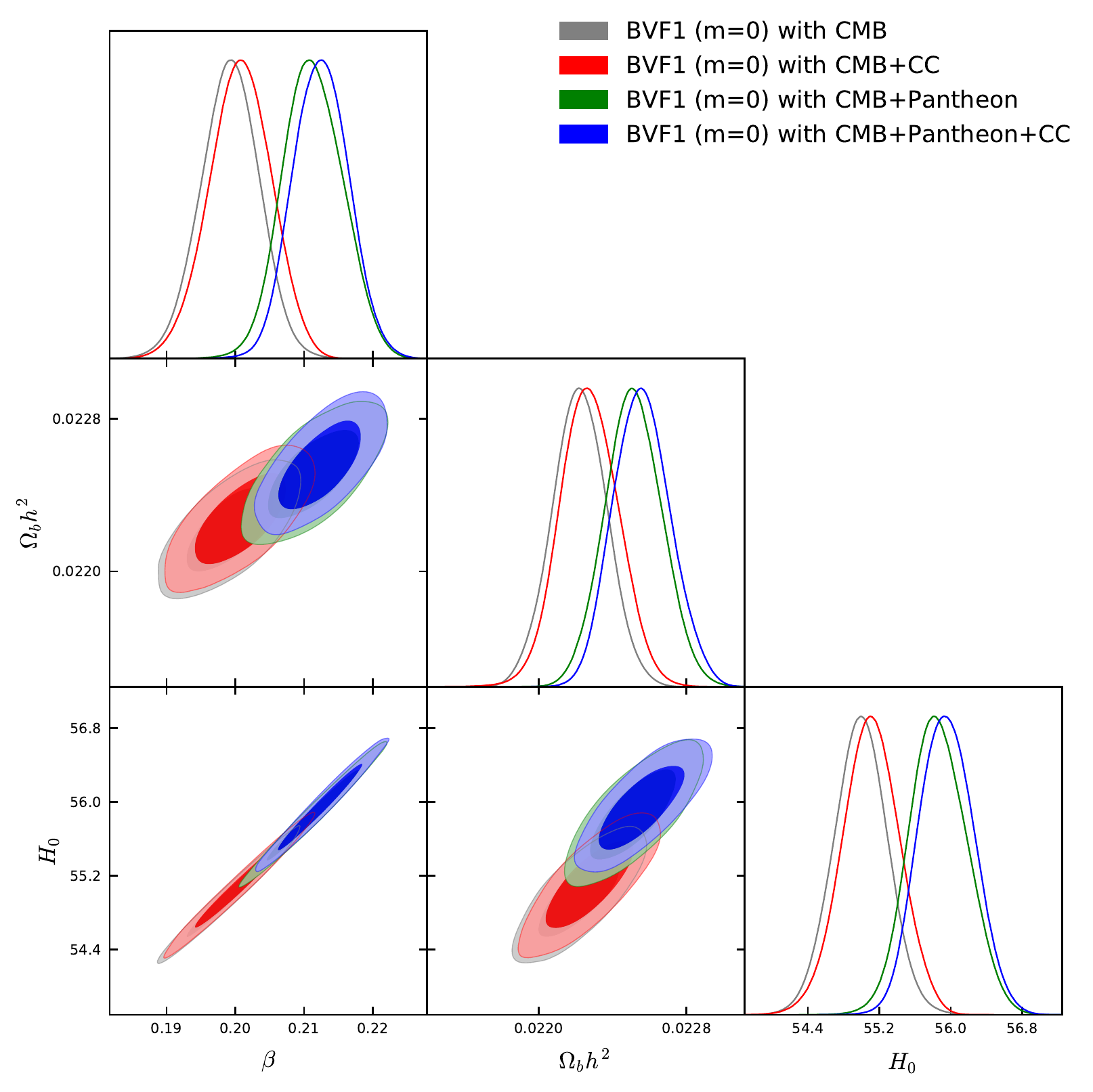}
\caption{68\% and 95\% c.l. contour plots for the BVF1 model with $m =0$, using the observational data from different sources. The figure also shows the one dimensional marginalized posterior distributions for some selected parameters.  }
\label{fig-bvf1-m0}
\end{figure*}

\begingroup                                                                                                                     
\squeezetable                                                                                                                   
\begin{center}                                                                                                                  
\begin{table*}                                                                                                                   
\begin{tabular}{cccccccccc}                                                                                                            
\hline\hline                                                                                                                    
Parameters & CMB & CMB+CC & CMB+Pantheon & CMB+Pantheon+CC\\ \hline

$\Omega_b h^2$ & $   0.02223_{-    0.00015-    0.00030}^{+    0.00015+    0.00030}$ &  $    0.02223_{-    0.00015-    0.00028}^{+    0.00015+    0.00030}$ & $    0.02219_{-    0.00015-    0.00031}^{+    0.00015+    0.00030}$ & $    0.02220_{-    0.00015-    0.00030}^{+    0.00015+    0.00031}$ \\

$100\theta_{MC}$ & $    1.0328_{-    0.0048-    0.0057}^{+    0.0046+    0.0052}$ &  $1.0293_{-    0.00097-    0.0021}^{+    0.00087+    0.0019}$ & $    1.02759_{-    0.00056-    0.0011}^{+    0.00057+    0.0011}$ & $    1.02759_{-    0.00056-    0.0011}^{+    0.00057+    0.0011}$ \\

$\tau$ & $    0.077_{-    0.017-    0.033}^{+    0.017+    0.031}$ & $ 0.079_{-    0.018-    0.036}^{+    0.018+    0.035}$ & $    0.077_{-    0.018-    0.033}^{+    0.017+    0.034}$  & $    0.078_{-    0.017-    0.034}^{+    0.017+    0.033}$ \\

$n_s$ & $    0.9646_{-    0.0047-    0.0092}^{+    0.0046+    0.0095}$ & $ 0.9652_{-    0.0047-    0.0092}^{+    0.0049+    0.0091}$ &  $    0.9652_{-    0.0046-    0.0089}^{+    0.0045 +    0.0091}$ & $    0.9653_{-    0.0045-    0.0088}^{+    0.0044+    0.0088}$ \\

${\rm{ln}}(10^{10} A_s)$ & $    3.089_{-    0.033-    0.064}^{+    0.033+    0.061}$ & $3.093_{-    0.034-    0.070}^{+    0.035+    0.068}$ & $    3.090_{-    0.034-    0.063}^{+    0.034+    0.066}$ & $    3.091_{-    0.034-    0.065}^{+    0.033+    0.065}$ \\

$\beta$ & $    0.22_{-    0.18-    0.20}^{+    0.20+    0.23}$ & $  0.365_{- 0.029-    0.066}^{+    0.038+    0.072}$ & $    0.428_{-    0.016-    0.034}^{+    0.016+    0.032}$ & $    0.429_{-    0.016-    0.033}^{+    0.017+    0.033}$ \\

$m$ & $   -0.06_{-    0.41-    0.51}^{+    0.40+    0.44}$ & $ -0.356_{-    0.088-    0.194}^{+    0.081+    0.176}$ & $   -0.544_{-    0.059-    0.13}^{+    0.066+    0.13}$ & $   -0.545_{-    0.058-    0.12}^{+    0.066+    0.12}$ \\

$H_0$ & $   56_{-   10-   12}^{+   11+   13}$ & $   64.1_{-    1.9-    3.9}^{+    2.0+    4.5}$ & $   68.0_{-    1.1-    2.3}^{+    1.1+    2.3}$ & $   68.1_{-    1.2-    2.2}^{+    1.1+    2.3}$ \\
\hline 
$\chi^2_{\rm best-fit}$ & 12957.848 & 12976.224 & 13997.658 & 14013.128 \\
\hline\hline                                                                                                                    
\end{tabular}                                                                                                                   
\caption{68\% and 95\% c.l. constraints on various free parameters of the model BVF1 with $m$ free using different observational data. Here $H_0$ is in the units of km/Mpc/sec.}
\label{tab:results-BVF1}                                                                                                   
\end{table*}                                                                                                                     
\end{center}                                                                                                                    
\endgroup                

\begin{figure*}
\includegraphics[width=0.7\textwidth]{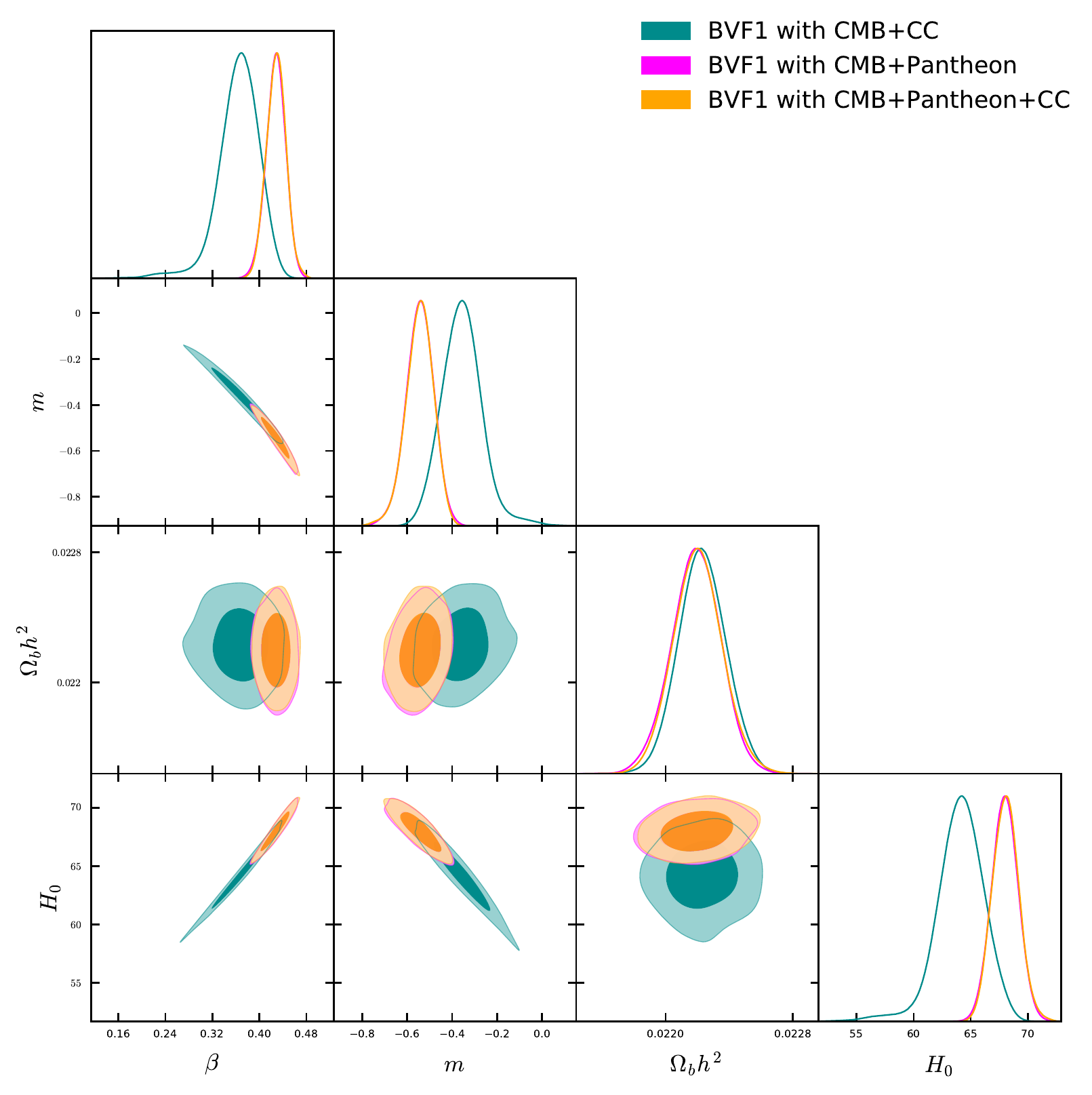}
\caption{The figure displays the 68\% and 95\% c.l. contour plots between various combinations of the parameters of the model BVF1 with $m$ free using the observational data from different sources. The figure also shows the one dimensional marginalized posterior distributions for some selected parameters. }
\label{fig-bvf1}
\end{figure*}
\begin{figure*}
\includegraphics[width=0.5\textwidth]{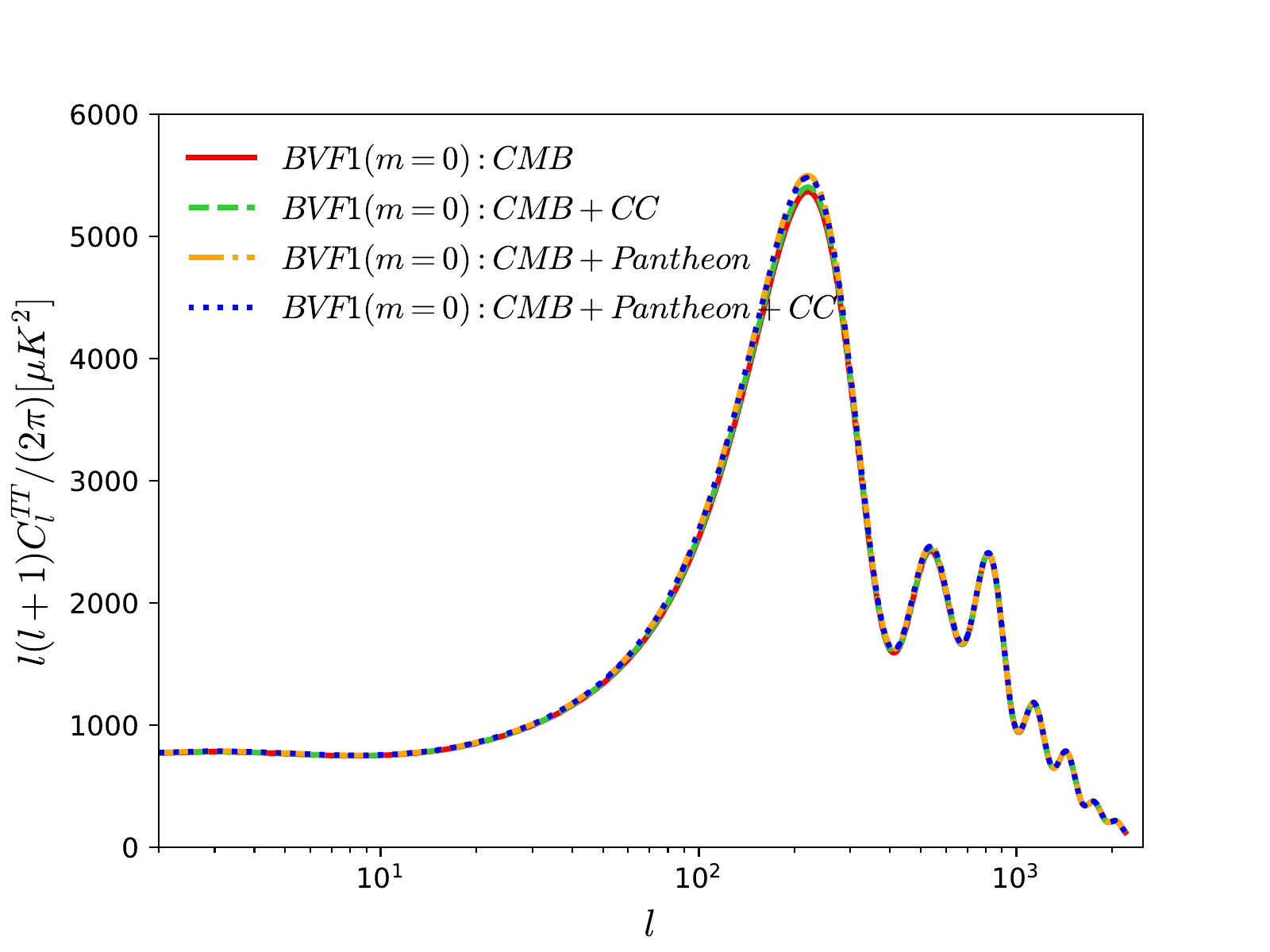}
\caption{CMB temperature power spectra  using the best-fit values of various free and derived parameters have been shown for BVF1 model with $m =0$ (i.e., the constant bulk viscosity). }
\label{fig-bvf1-cmbmp0}
\end{figure*}
\begin{figure*}
\includegraphics[width=0.5\textwidth]{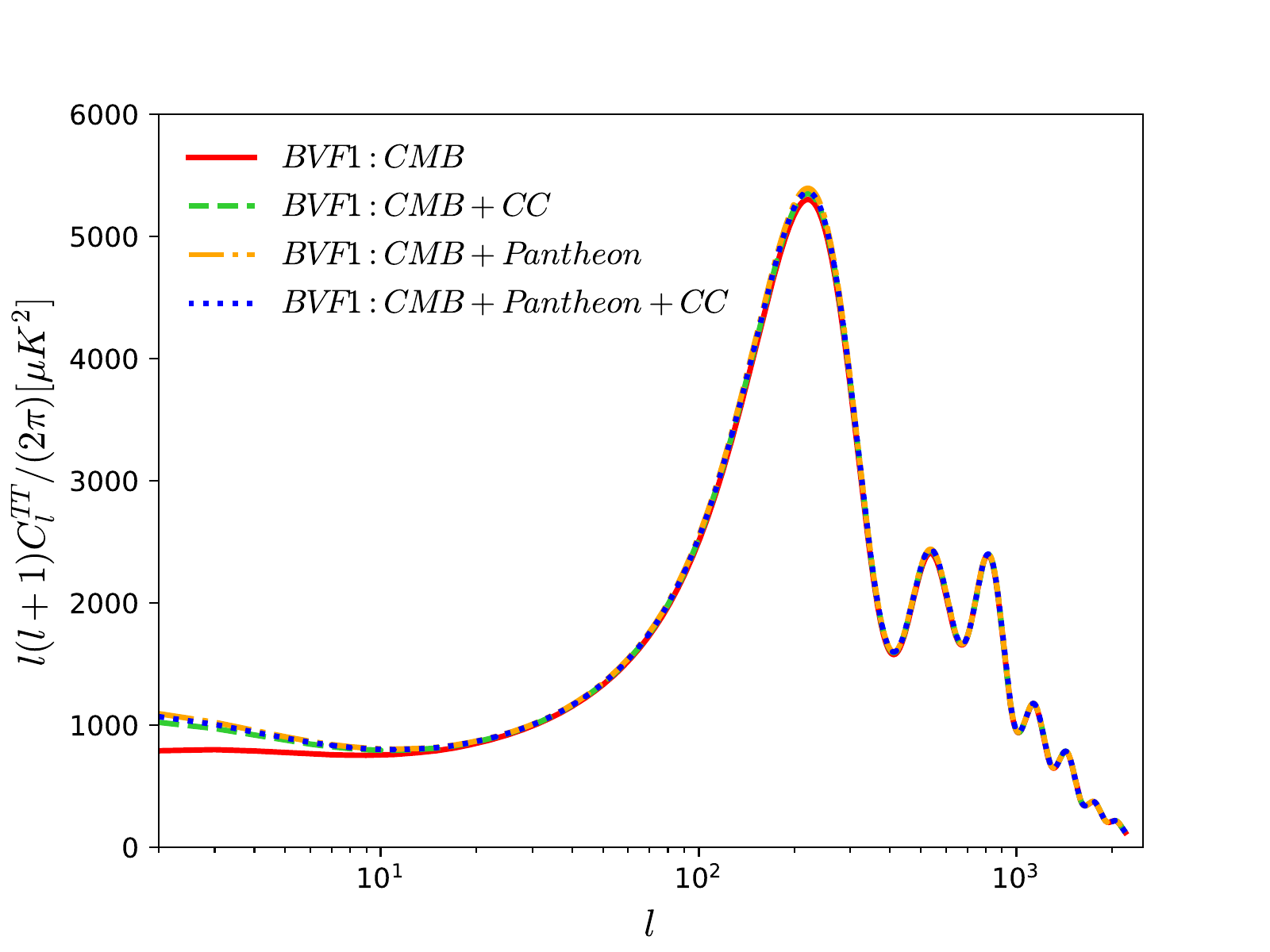}
\caption{CMB temperature power spectra using the best-fit values of various free and derived parameters of the generalized the bulk viscous model BVF1. }
\label{fig-bvf1-cmbmp_general}
\end{figure*}

\subsection{The model BVF1}
\label{sec-bvf1}

Here we present the observational summary of the BVF1 model using various combinations of the cosmological datasets. The governing parameters of this model as already mentioned are $m$ and $\beta$ (we recall again that $\beta = \alpha\rho_{t0}^{m-1/2}$).  We first consider the simplest bulk viscous scenario with $m =0$ that represents a constant bulk viscosity in the universe sector and then proceed towards the more general scenario where $m$ has been taken to be a free parameter.  

For the scenario with $m = 0$, we have constrained the model using four different observational datasets, namely, CMB, CMB+CC, CMB+Pantheon, and CMB+Pantheon+CC, and the results of the scenario are summarized in Table \ref{tab:bvf1-m0}. From Table \ref{tab:bvf1-m0}, one may notice that the results can be divided in two different blocks, with and without Pantheon, while they are about insensitive to the presence of CC. In particular, we see that adding Pantheon we have a shift of all the cosmological parameters, except $\theta_{MC}$, towards higher values. Moreover, one can see that here for $\theta_{MC}$ and $H_0$ the CMB case goes down several standard deviations, about~$20$, compared to the Planck's $\Lambda$CDM based estimation \cite{Ade:2015xua}. This is a very striking result since the estimated values of $\theta_{MC}$ and $H_0$ for $m =0$ have large difference to that of the Planck \cite{Ade:2015xua}, and the $H_0$ constraint is twice stronger of the $\Lambda$CDM one. On the other hand, the constant $\beta$ assumes small values and it is always different from zero at more than $4$ standard deviations, which goes in favor of the bulk viscosity. In Fig. \ref{fig-bvf1-m0} we have shown the one dimensional marginalized posterior distributions for some free parameters as well as the two dimensional contour plots considering various dataset combinations for the BVF1 model. From this figure (i.e., Fig. \ref{fig-bvf1-m0}), one can clearly see that the parameters are correlated with one another. In particular, there is a strong positive correlation amongst the parameters shown in the figure for this model.

Now, concerning the general scenario where $m$ acts as a free parameter, we have summarized the observational constraints on the model parameters in Table \ref{tab:results-BVF1} at 68\% and 95\% CL. From our analyses (see Table \ref{tab:results-BVF1}), we see that the CMB data alone return very low value of the Hubble parameter at present, i.e., $H_0 = 56_{- 10}^{+ 11}$ at 68\% CL but with large error bars, and of $\theta_{MC}$, i.e., $100\theta_{MC} = 1.0328^{+0.0046}_{-0.0048}$. Subsequently, when the external data sets are added to CMB, the error bars on $H_0$ significantly decrease and $H_0$ increases. We see that the addition of CC to CMB gives better constraints on $H_0$ and $\theta_{MC}$, on the contrary to the results with respect to the $m=0$ case. However, the best constraints are achieved for the addition of the Pantheon data to the CMB, and the results for this dataset (i.e., CMB+Pantheon) are practically indistinguishable from the full combination CMB+Pantheon+CC. Thus, in order to show the graphical variations for the model parameters, we limit to three combined analyses, namely, CMB+CC, CMB+Pantheon, and CMB+Pantheon+CC, because the CMB only constraints are too much large and the figure should be unreadable if added. In Fig.~\ref{fig-bvf1}, for the above three datasets, we show the 1D marginalized posterior distributions for the free parameters of the model as well as the 2D contour plots for several combinations of the free parameters at 68\% and 95\% CL. From this plot we can see the strong anticorrelation between $m$ and $\beta$.  
Moreover, from Fig. \ref{fig-bvf1} we have some common features of some parameters that are independent of the datasets.  We see that the parameter $\beta$ has a strong positive correlation to $H_0$ and this is independent of the datasets used. Correspondingly, we find that the parameter $m$ has a strong negative correlation to $H_0$. 
Further, one can notice that there is a notable shift between the constraints from CMB, CMB+CC and CMB+Pantheon. For the dataset CMB+CC, the estimation of $H_0$ ($= 63.2_{-    1.6}^{+    2.4}$, 68\% CL) moves towards a higher value with respect to CMB alone, but is still slightly far from the measurements by Planck~\cite{Ade:2015xua} in the $\Lambda$CDM scenario, while for the dataset CMB+Pantheon, the estimated value of  $H_0$ ($= 68.0\pm 1.1$, 68\% CL) is similar to Planck  \cite{Ade:2015xua} but with slightly large error bars. 
{\it Interestingly, one can notice that due to large error bars on $H_0$ for this dataset (i.e., CMB+Pantheon), it is possible to weaken the tension on $H_0$ observed from the local estimation of $H_0$ measured by Riess et al. in 2016: $H_0 = 73.24 \pm 1.74$ km $s^{-1}$ Mpc$^{-1}$ \cite{Riess:2016jrr}  and in 2018: $H_0 = 73.48 \pm 1.66$ km $s^{-1}$ Mpc$^{-1}$ \cite{Riess:2018uxu} under $3\sigma$ CL.} However, if we consider the updated value of the present day Hubble constant, $H_0 = 74.03 \pm 1.42$ km $s^{-1}$ Mpc$^{-1}$ of Riess et al. 2019 \cite{Riess:2019cxk} the tension is still at $3.3\sigma$. This is one of the interesting results in this context since the $H_0$ tension is partially alleviated, even if this is probably due to a volume effect, i.e. to the large error bars imposed by the observational data.
On the contrary, with respect to the case with $m=0$ the $\theta_{MC}$ parameter shifts towards lower values moving away from Planck~\cite{Ade:2015xua} in the $\Lambda$CDM scenario when adding CC or Pantheon to CMB.
Finally, for the last combination, that means, CMB+Pantheon+CC, we find identical constraints compared to CMB+Pantheon, showing that CC is not adding any new information to the analysis.  A similar observation can be found from the constraints on the model parameters, specifically looking at the constraints on $m$ and $\beta$,  we can see that the combinations CMB+CC, CMB+Pantheon and CMB+Pantheon+CC significantly improve the parameters space compared to the constraints obtained only from the CMB data alone.   
In fact, while for the CMB case we have an indication at two standard deviations for $\beta$ greater than zero, this becomes a very robust evidence at several standard deviations after the inclusion of other cosmological probes. In particular, we have a shift towards higher values, passing from $\beta=0.22^{+0.20}_{-0.18}$ at 68\% CL for CMB alone, to $\beta=0.350^{+0.042}_{-0.026}$ at 68\% CL for CMB+CC, to $\beta=0.428\pm 0.016$ at 68\% CL for CMB+Pantheon. Analogously, we see a shift of the $m$ parameter towards lower values, due to the negative correlation with $\beta$ leaving with a solid evidence for $m$ to be negative and different form zero at about roughly $8$ standard deviations for the full combination CMB+Pantheon+CC. Also for this parameter, while the CMB alone value is in agreement with $m=0$ ($m=-0.06^{+0.40}_{-0.41}$ at 68\% CL), it moves to $m=-0.320^{+0.073}_{-0.099}$ at 68\% CL after the inclusion of CC, to $m=-0.544^{+0.066}_{-0.059}$ at 68\% CL for CMB+Pantheon. If we compare the Table \ref{tab:bvf1-m0} with $m=0$ with the Table \ref{tab:results-BVF1} with $m$ free, we see an exceptional gain of $\Delta \chi^2=150$, supporting the necessity of $m$ different from zero.

\subsubsection{The BVF1 model at large scales}
\label{subsec-bvf1-largescale}

We now discuss the behaviour of the BVF1 model at the level of perturbations considering the direct impacts on the CMB TT spectra. We start with the constant bulk viscous scenario (i.e., $m = 0$) and display the CMB TT spectra  in  Fig.  \ref{fig-bvf1-cmbmp0}, using the best-fit values of all model parameters obtained for different observational datasets. From different spectrum corresponding to each observational data we cannot distinguish them.

For the second scenario of this model with free $m$ (i.e., model BVF2), we perform similar investigation as with BVF1 ($m =0$). In Fig. \ref{fig-bvf1-cmbmp_general}, we show the CMB TT spectra for all the observational datasets using the best-fit values of the model parameters. We find that a difference in the spectra between CMB and the remaining observational datasets in the lower multipoles (for $l \leq 10$), i.e. in the cosmic variance limited region, while for higher multipoles (for $l > 10$), we do not observe any changes in the curves.

\begingroup                                                                                                                     
\squeezetable                                                                                                                   
\begin{center}                                                                                                                  
\begin{table*}                                                                                                                   
\begin{tabular}{ccccccccccccc}                                                                                                            
\hline\hline                                                                                                                    
Parameters & CMB & CMB+CC & CMB+Pantheon  & CMB+Pantheon+CC\\ \hline

$\Omega_b h^2$ & $    0.02221_{-    0.00019-    0.00033}^{+    0.00017+    0.00034}$ & $    0.02209_{-    0.00015-    0.00030}^{+    0.00016+    0.00030}$ & $    0.02196_{-    0.00017-    0.00031}^{+    0.00016+    0.00032}$ & $    0.02199_{-    0.00016-    0.00033}^{+    0.00016+    0.00031}$ \\

$100\theta_{MC}$ & $    1.0328_{-    0.0014-    0.0029}^{+    0.0015+    0.0027}$ & $    1.03081_{-    0.00089-    0.0016}^{+    0.00071+    0.0017}$ & $    1.02676_{-    0.00073-    0.0018}^{+    0.00092+    0.0015}$ & $    1.02759_{-    0.00075-    0.0014}^{+    0.00070+    0.0014}$ \\

$\tau$ & $    0.079_{-    0.017-    0.034}^{+    0.017+    0.034}$ & $    0.086_{-    0.018-    0.033}^{+    0.017+    0.034}$ & $    0.104_{-    0.019-    0.035}^{+    0.019+    0.037}$ & $    0.101_{-    0.018-    0.036}^{+    0.019+    0.035}$ \\

$n_s$ & $    0.9668_{-    0.0083-    0.018}^{+    0.0086+    0.017}$ & $    0.9774_{-    0.0053-    0.011}^{+    0.0062+    0.010}$ & $    0.9995_{-    0.0057-    0.011}^{+    0.0054+    0.012}$ & $    0.9953_{-    0.0055-    0.010}^{+    0.0054+    0.010}$ \\

${\rm{ln}}(10^{10} A_s)$ & $    3.095_{-    0.035-    0.068}^{+    0.035+    0.068}$ & $    3.112_{-    0.036-    0.065}^{+    0.034+    0.066}$  & $    3.154_{-    0.037-    0.070}^{+    0.037+    0.073}$ & $    3.147_{-    0.034-    0.069}^{+    0.037+    0.068}$ \\

$\beta$ & $    0.209_{-    0.035-    0.077}^{+    0.045+    0.075}$ & $    0.261_{-    0.012-    0.034}^{+    0.019+    0.029}$ & $    0.333_{-    0.012-    0.022}^{+    0.011+    0.022}$ & $    0.320_{-    0.009-    0.020}^{+    0.011+    0.020}$ \\

$\gamma$ & $    1.001_{-    0.003-    0.006}^{+    0.003+    0.006}$ & $    1.005_{-    0.001-    0.003}^{+    0.002+    0.003}$ & $    1.013_{-    0.002-    0.003}^{+    0.001+    0.003}$ & $    1.012_{-    0.001-    0.002}^{+    0.001+    0.002}$ \\

$H_0$ & $   56.1_{-    3.6-    6.1}^{+    3.4+    6.3}$ & $   60.8_{-    1.4-    3.5}^{+    1.9+    2.9}$ & $   70.2_{-    1.9-    3.3}^{+    1.6+    3.7}$ &  $   68.2_{-    1.4-    2.8}^{+    1.4+    2.9}$\\

\hline 

$\chi^2_{\rm best-fit}$ & 12962.760 & 12981.476 & 14037.120  & 14059.740 \\

\hline\hline                                                                                                                    
\end{tabular}                                                                                                                   
\caption{68\% and 95\% c.l. constraints on various free parameters of the model BVF2 assuming the simplest case $m =0$, that means, the constant bulk viscosity, using different observational data. Here $H_0$ is in the units of km/Mpc/sec.}
\label{tab:results-BVF2-m0}                                                                                                   
\end{table*}                                                                                                                     
\end{center}                                                                                                                    
\endgroup                                                                                                                       

\begin{figure*}
\includegraphics[width=0.7\textwidth]{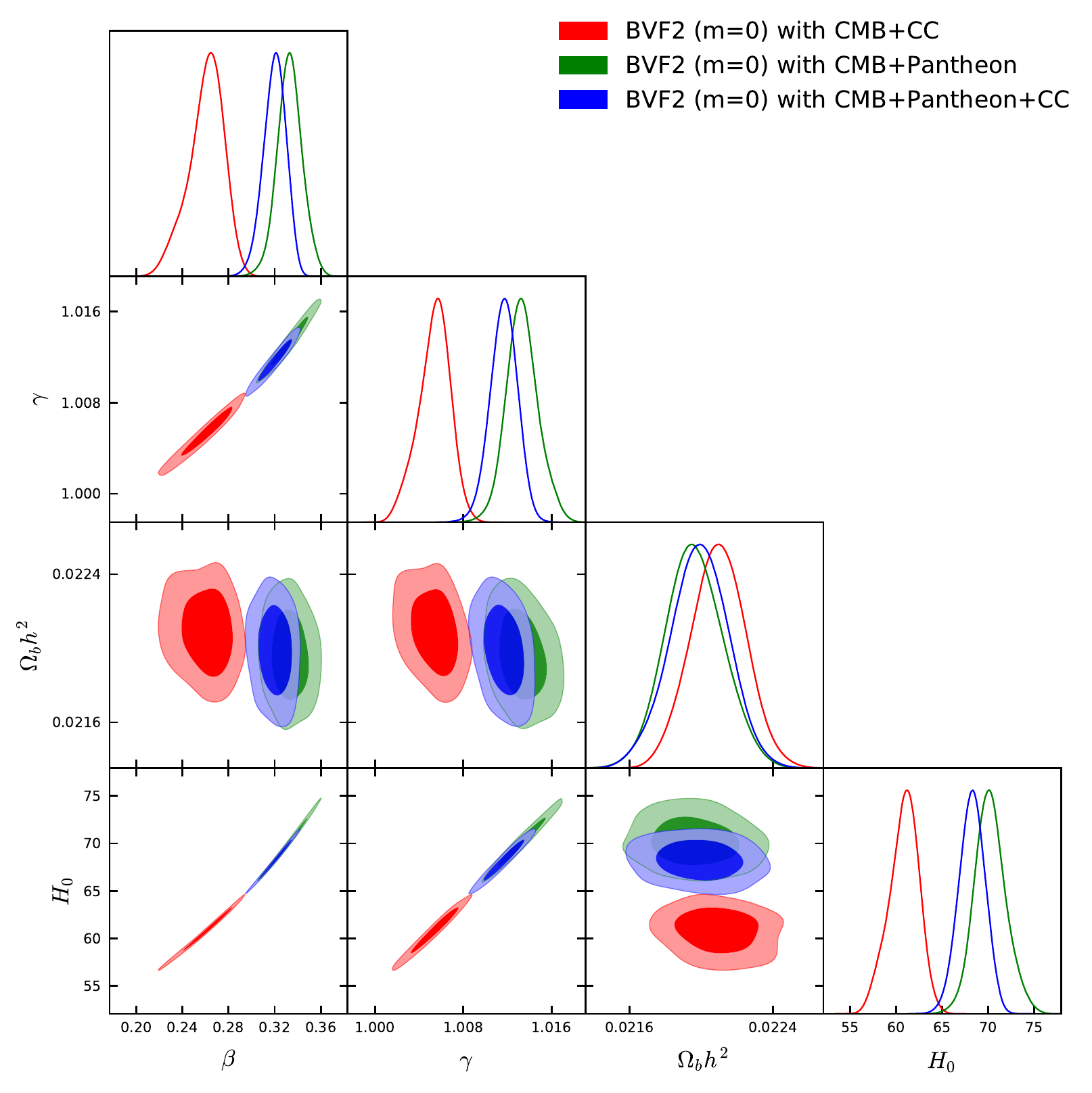}
\caption{68\% and 95\% c.l. contour plots for the BVF2 model with $m =0$, using the observational data from different sources. The figure also shows the one dimensional marginalized posterior distributions for some selected parameters. }
\label{fig-bvf2-m0}
\end{figure*}

\begingroup                                                                                                                     
\squeezetable                                                                                                                   
\begin{center}                                                                                                                  
\begin{table*}                                                                                                                   
\begin{tabular}{ccccccc}                                                                                                            
\hline\hline                                                                                                                    
Parameters & CMB & CMB+CC & CMB+Pantheon  & CMB+Pantheon+CC\\ \hline
$\Omega_b h^2$ & $    0.02220_{-    0.00018-    0.00033}^{+    0.00018 +    0.00034}$ & $ 0.02218_{-    0.00016-    0.00031}^{+    0.00016+    0.00032}$  & $    0.02219_{-    0.00017-    0.00032}^{+    0.00016+    0.00033}$ & $    0.02219_{-    0.00016-    0.00032}^{+    0.00016+    0.00032}$ \\

$100\theta_{MC}$ & $    1.0330_{-    0.0040-    0.0051}^{+    0.0040+    0.0048}$ & $ 1.0284_{-    0.0014-    0.0029}^{+    0.0017+    0.0026}$ & $    1.0276_{-    0.0012-    0.0020}^{+    0.0011+    0.0021}$ &  $    1.02741_{-    0.00083-    0.0015}^{+    0.00077+    0.0016}$  \\

$\tau$ & $    0.078_{-    0.019-    0.039}^{+    0.021+    0.037}$ & $ 0.080_{-    0.019-    0.034}^{+    0.018+    0.036}$  & $    0.079_{-    0.017-    0.033}^{+    0.017+    0.033}$ & $    0.078_{-    0.017-    0.033}^{+    0.017+    0.033}$ \\

$n_s$ & $    0.9670_{-    0.0079-    0.014}^{+    0.0072+    0.015}$ & $ 0.9690_{-    0.0062-    0.0125}^{+    0.0061+    0.0129}$ & $    0.9659_{-    0.0081-    0.0134}^{+    0.0081+    0.014}$ & $    0.9668_{-    0.0054-    0.011}^{+    0.0055+    0.011}$ \\

${\rm{ln}}(10^{10} A_s)$ & $    3.091_{-    0.037-    0.074}^{+    0.042+    0.074}$ &  $  3.097_{-    0.035-    0.068}^{+    0.036+    0.068}$ & $    3.093_{-    0.033-    0.067}^{+    0.034+    0.066}$ & $    3.092_{-    0.033-    0.063}^{+    0.034 +    0.065}$ \\

$\beta$ & $    0.21_{-    0.15-    0.17}^{+    0.15+    0.19}$ & $0.378_{-    0.050-    0.10}^{+    0.065+    0.097}$ & $    0.427_{-    0.024-    0.038}^{+    0.020+    0.041}$ & $    0.424_{-    0.017-    0.033}^{+    0.018+    0.035}$ \\

$m$ & $    0.00_{-    0.17-    0.43}^{+    0.29+    0.37}$ & $-0.35_{-0.18-    0.31}^{+    0.21+    0.31}$ & $   -0.56_{-    0.12-    0.41}^{+    0.25+    0.31}$ & $   -0.51_{-    0.09-    0.23}^{+    0.14+    0.21}$ \\

$\gamma$ & $    1.001_{-    0.002-    0.005}^{+    0.002+    0.005}$ & $1.002_{-    0.002-    0.004}^{+    0.002+    0.004}$ & $    1.000_{-    0.003-    0.005}^{+    0.003+    0.005}$ & $    1.001_{-    0.002-    0.003}^{+    0.002+    0.003}$ \\

$H_0$ & $   55.8_{-    9.1-   10.9}^{+    9.2+   11.6}$ & $ 66.2_{-    3.8-    5.6}^{+    3.1+    6.2}$ & $   68.0_{-    2.4-    4.7}^{+    2.7+    4.5}$ & $   68.4_{-    1.5-    3.3}^{+    1.8 +    3.0}$ \\
\hline 
$\chi^2_{\rm best-fit}$ & 12959.084 & 12977.172 &  13998.418 &  14013.918 \\
\hline\hline                                                                                                                    
\end{tabular}                                                                                                                   
\caption{68\% and 95\% c.l. constraints on various free parameters of the model BVF2 using different observational data. Here $H_0$ is in the units of km/Mpc/sec. }
\label{tab:results-BVF2}                                                                                                   
\end{table*}                                                                                                                     
\end{center}                                                                                                                    
\endgroup                                                                                                                       

\begin{figure*}
\includegraphics[width=0.7\textwidth]{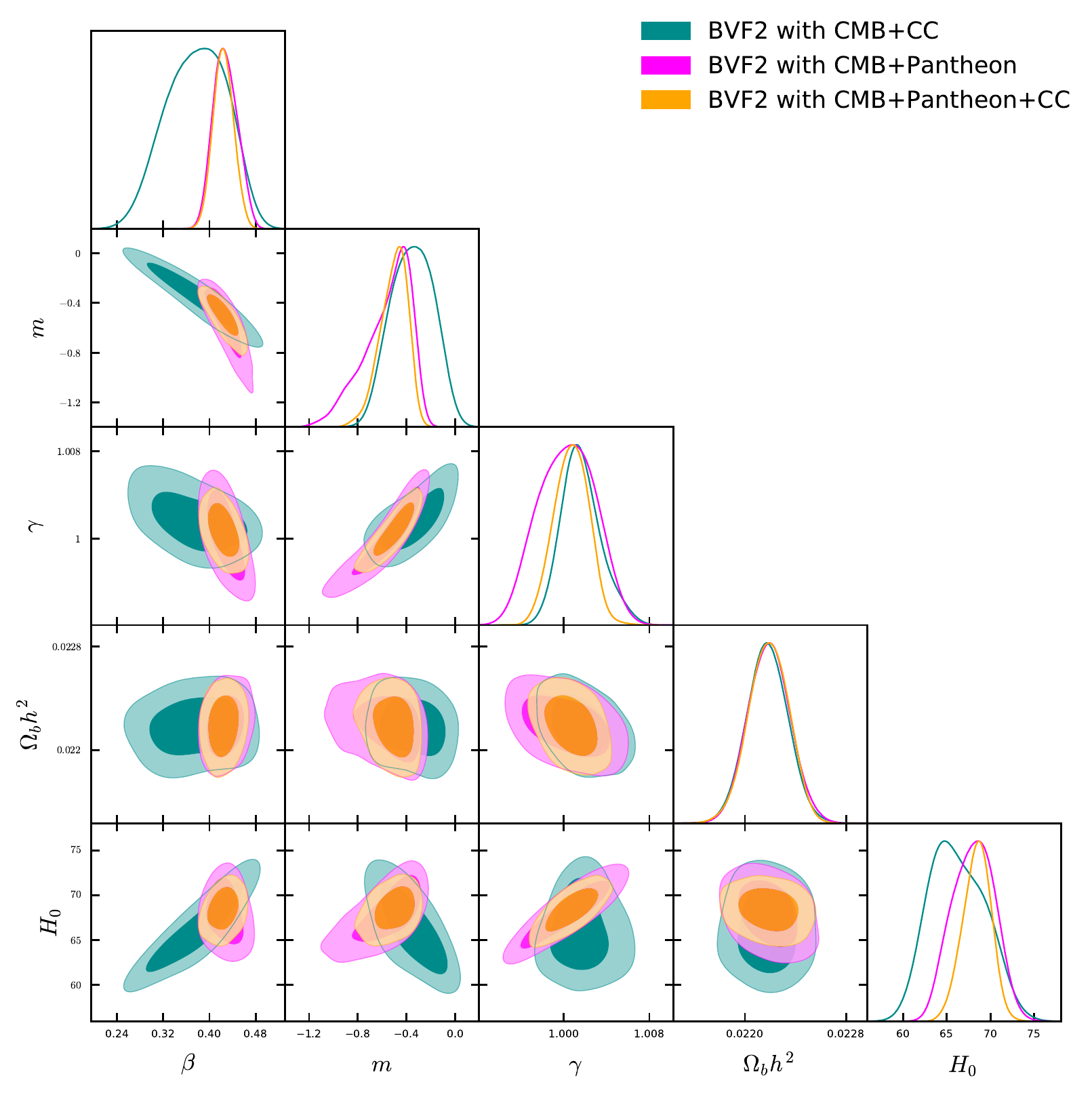}
\caption{The figure displays the 68\% and 95\% c.l. contour plots between various combinations of the parameters of the model BVF2 using the observational data from different sources. The figure also shows the one dimensional marginalized posterior distributions for some selected parameters.  }
\label{fig-bvf2}
\end{figure*}
\begin{figure*}
\includegraphics[width=0.5\textwidth]{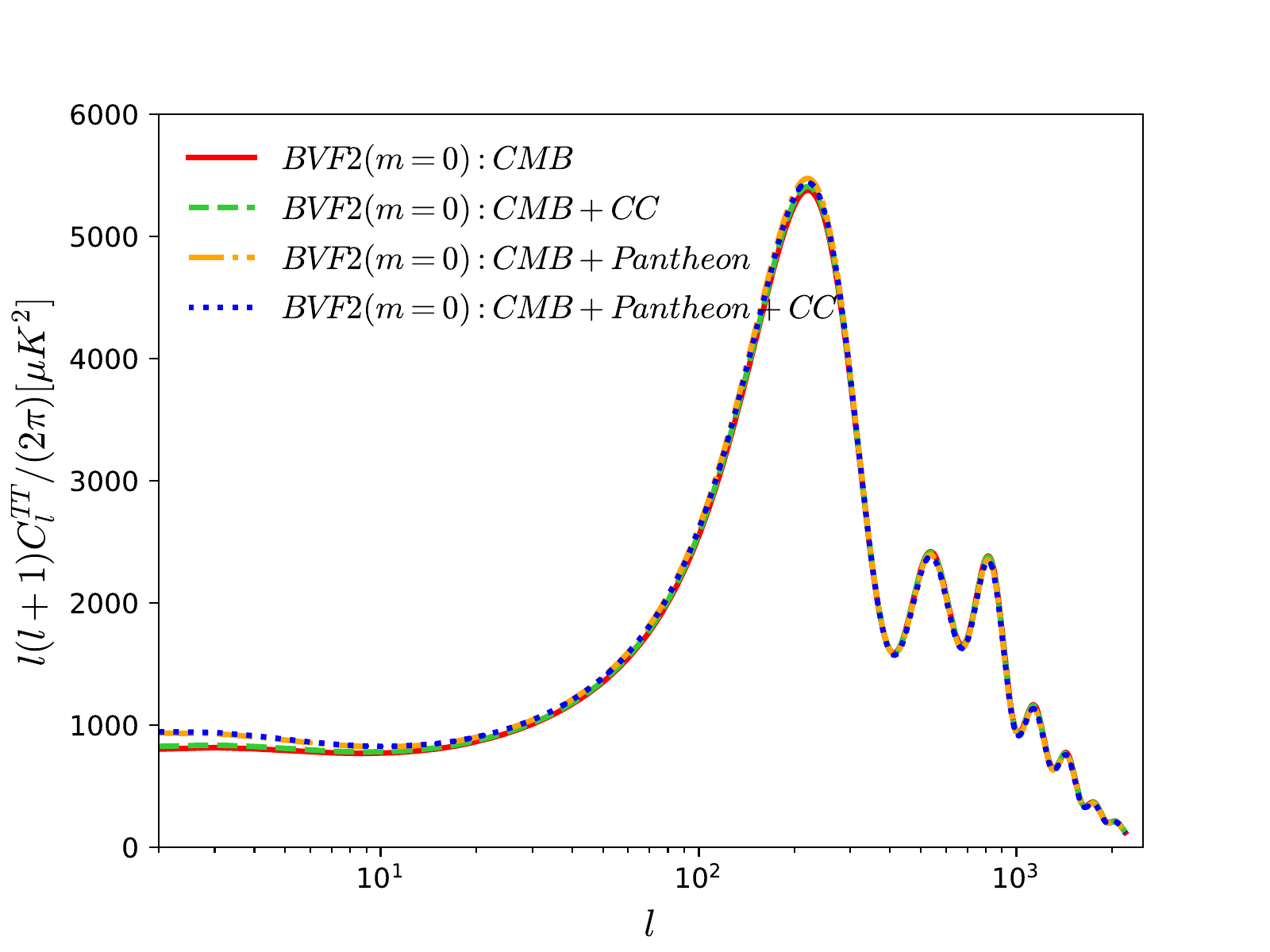}
\caption{We show the CMB TT spectra for the BVF2 model with $m=0$ using the best-fit values of the free and derived parameters. }
\label{fig-bvf2-cmbm0}
\end{figure*}
\begin{figure*}
\includegraphics[width=0.5\textwidth]{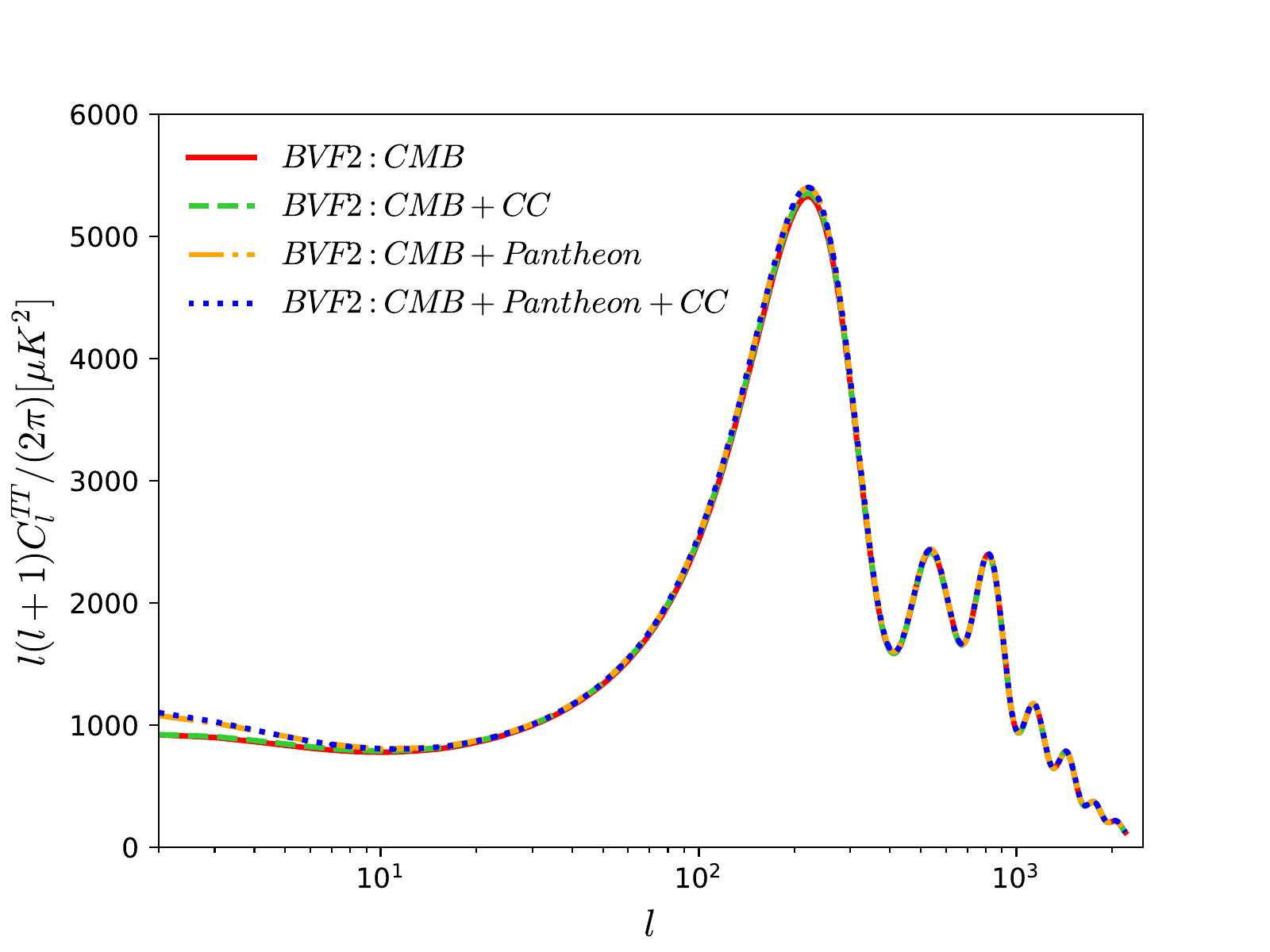}
\caption{Temperature anisotropy in the CMB spectra  for different best-fit values of the free and derived parameters of the generalized bulk viscous scenario BVF2. }
\label{fig-bvf2-cmbmp_general}
\end{figure*}

\subsection{The model BVF2}
\label{sec-bvf2}

The main free parameters of this model are $\beta$, $\gamma$ and $m$. So, following a similar fashion as performed with model BVF1, here too, we consider the constant bulk viscosity scenario leading to $m =0$, as the simplest possibility and then proceed towards the more general bulk viscous scenario assuming $m$ to be  a free parameter.  
   
In Table \ref{tab:results-BVF2-m0}, we show the observational constraints for the constant bulk viscous model (the $m =0$ case) at the 68\% and 95\% CL. From the Table \ref{tab:results-BVF2-m0}, one can readily see that
the error bars are strongly relaxed, until an order of magnitude, with respect to the BVF1 case. We also find that in this BVF2 scenario a large shift of $\theta_{MC}$ at about $5$ standard deviations and $H_0$ towards lower values, with respect to the $\Lambda$CDM model, but thanks to the larger error bars, now the shift is less significant.
In particular, we have $H_0 = 56.1_{-    3.6}^{+    3.4}$ km/s/Mpc at 68\% CL for the dataset CMB only, to be compared to its estimation by Planck  \cite{Ade:2015xua} $H_0 = 67.27 \pm 0.66$ km/s/Mpc at 68\% CL.
When the external datasets, such as CC and Pantheon are added to CMB, $H_0$ goes up (more in agreement with the $\Lambda$CDM value) and $\theta_{MC}$ goes down (increasing the disagreement with the $\Lambda$CDM value), and the error bars on them are also reduced compared to their estimation from CMB only.
Moreover, $\gamma$ shifts away from $1$ (that corresponds to the BVF1 scenario) of several standard deviations when more datasets are combined together.
For this model, contrary to the BVF1 case, the addition of the CC dataset improves the parameter constraints, although for the CMB+CC data the estimated value of $H_0= 60.8_{- 1.4}^{+ 1.9}$ km/s/Mpc at 68\% CL, is still lower than  \cite{Ade:2015xua}.
When considering the Pantheon datasets and the full combination, namely CMB+Pantheon and CMB+Pantheon+CC, the Hubble constant $H_0$ has the strongest constraints and it shifts towards higher values. {\it Therefore, it is quite interesting to notice that for the CMB+Pantheon dataset, the tension in $H_0$ with the local measurements is clearly reconciled within 95\% CL, even considering the latest Riess et al. 2019 measurement \cite{Riess:2019cxk}.
For the full combination CMB+Pantheon+CC, the tension is also released but within $3\sigma$. } 
Concerning the $\beta$ parameter, we always find that it is different from zero at more than 95\% CL, irrespective of the observational datasets. We again note that for the CMB only data, $\beta$ has the maximum error bars which are eventually decreased after the addition of external datasets such as CC or Pantheon or both, and moreover, we further note that the strongest constraint on $\beta$ is achieved for the full combination CMB+Pantheon+CC.  
To compare the different datasets, we choose the last three, namely CMB+CC, CMB+Pantheon, CMB+Pantheon+CC and in Fig.~\ref{fig-bvf2-m0} we show the one dimensional marginalized posterior distributions for the model parameters as well the contour plots between several combinations of the same free parameters at 68\% and 95\% CL. The Fig. \ref{fig-bvf2-m0} offers some interesting behaviour of the parameters. First of all, we can notice also in this case that the addition of Pantheon produces a large shift of the cosmological parameters, now in tension with the estimates obtained by the CMB+CC dataset combination. This shift indicates a disagreement of the Pantheon dataset with the CMB in the context of the BVF model. Secondly, we find that the parameters $H_0$, $\beta$ and $\gamma$ have strong positive correlations between them, while the contours of ($\beta$, $\Omega_{b}h^2$) and ($\gamma$, $\Omega_{b}h^2$) are almost vertical leading to no correlations between them.   

Now, concerning the general bulk viscous scenario with free $m$, we have analyzed it using the same observational datasets such as CMB, CMB+CC, CMB+Pantheon and CMB+Pantheon+CC and we present the observational constraints on the model parameters in Table \ref{tab:results-BVF2}. From the analyses (referred to Table \ref{tab:results-BVF2}), one can see that for the CMB data alone, $H_0$ confirms a lower mean  value like in the $m=0$ case, but with higher error bars, $H_0 = 55.8_{-    9.1}^{+    9.2}$ (68\% CL, CMB), and again, when the external datasets are added, namely the CMB+CC or CMB+Pantheon or CMB+Pantheon+CC, the error bars are significantly reduced with increased values of the Hubble constant in this way: $H_0 =65.9_{-4.0}^{+  3.5} $ (68\% CL, CMB+CC); $H_0 = 68.0_{- 2.4}^{+ 2.7}$ (68\% CL, CMB+Pantheon) and $H_0 = 68.4_{-1.5}^{+1.8}$ (68\% CL, CMB+Pantheon+CC). 
While one can recognize that the estimated value of $H_0$ for the combination CMB+CC, is still lower than the Planck's estimation \cite{Ade:2015xua}, interestingly, 
for the last two datasets, namely CMB+Pantheon and CMB+Pantheon+CC, we see that due to larger error bars (which are indeed very small attained for the analysis with CMB alone), the estimated values of $H_0$ can still catch the local estimation of $H_0$ by Riess et al. 2019 \cite{Riess:2019cxk} within $3\sigma$. {\it Eventually, the tension on $H_0$ is clearly reduced, which is indeed one of the most interesting properties of this bulk viscous model.  }

Let us now focus on the constraints of other free parameters. From the constraints on $\gamma$ (see Table \ref{tab:results-BVF2}), we see that, this parameter is very very close  and in agreement to $1$. The numerical estimations of $\gamma$ from different observational datasets do not change much from one to another:
$ \gamma = 1.001_{- 0.002}^{+    0.002}$ (68\% CL, CMB); $ \gamma =  1.002_{- 0.002}^{+    0.001}$ (68\% CL, CMB+CC); $ \gamma = 1.000_{-    0.003}^{+    0.003}$ (68\% CL, CMB+Pantheon); and $\gamma = 1.001_{-    0.002}^{+    0.002}$ (68\% CL, CMB+Pantheon+CC).   
However, a significant changes appear in the constraints of the $m$ parameter (let us recall that the $m$ parameter appears from the bulk viscosity: $\eta (\rho_D) = \alpha \rho_D^{m}$, $\alpha > 0$). From the CMB data alone, we see that $m = 0.00_{-    0.17}^{+    0.29}$ (68\% CL) whilst from the remaining datasets, the mean values of $m$ are negative with increased significance when more datasets are considered: $m = -0.32_{-    0.21}^{+    0.16}$ (68\% CL, CMB+CC), $m = -0.56_{-    0.12}^{+    0.25} $ (68\% CL, CMB+Pantheon), $ m = -0.51_{-0.09}^{+    0.14}$ (68\% CL, CMB+Pantheon+CC).  From all the analyses, we find that $m \neq 0$ at more than 68\% CL, that means, it gives a strong indication for a bulk viscous scenario apart from the constant one.

Finally, we discuss the observational bounds on $\alpha$ in terms of the $\beta$ parameter quantifying the bulk viscosity in the universe sector. As already reported, the best constraints for $\beta$ is achieved for the dataset CMB+Pantheon+CC with $\beta  = 0.424_{- 0.017}^{+ 0.018}$ at 68\% CL. Thus, overall we find that the observational data are in support of  a bulk viscous cosmology.  We close this section with Fig. 
Fig. \ref{fig-bvf2}, where for the last three best analyses, namely, CMB+CC, CMB+Pantheon and CMB+Pantheon+CC, 
we display the one dimensional marginalized posterior distributions for the free parameters of the model as well as the contour plots between various combinations of the model parameters at 68\% and 95\% CL. From this figure (i.e., Fig. \ref{fig-bvf2}), we clearly see that the parameter $m$ has a strong positive correlation to  $\gamma$ and $\beta$ has a negative correlation to both $m$ and $\gamma$, partially broken by the addition of the Pantheon dataset. Moreover, the parameter $m$ presents a positive correlation with $H_0$ for the CMB+CC case, while the addition of the Pantheon dataset changes the direction of the correlation. This is the reason why by adding the Pantheon dataset the $H_0$ value is very well constrained, shifting $m$ towards negative values. 
In Fig. \ref{fig-bvf2} all the bounds are now very well consistent, therefore we can conclude that having a negative $m$ parameter is a way to solve the disagreement between the CMB and Pantheon datasets we saw in Fig. \ref{fig-bvf2-m0}. 
The full combination of datasets considered in this work is therefore converging to a concordance model with a negative $m$ at several standard deviations, a larger $\beta$ different from zero, a $\gamma$ consistent with $1$, a larger value for the Hubble constant and a smaller value for $\theta_{MC}$. For this model we gain a $\Delta \chi^2=46$ with respect to the same case 
with $m=0$.

\subsubsection{The BVF2 model at large scales}
\label{subsec-bvf2-largescale}

We continue by discussing the effects on the CMB TT and matter power spectra for the two variations of this bulk viscous scenario, namely for $m =0$ and with free $m$. 

As far as the simplest case with constant bulk viscous model (i.e., $m =0$) is concerned, in Fig. \ref{fig-bvf2-cmbm0} we plot the CMB TT spectra. To draw the plot we have used the best-fit values of the model parameters obtained from all the observational datasets that we have used in this work. From Fig. \ref{fig-bvf2-cmbm0}, we find that at lower multipoles (i.e., $l \leq 10$), a mild deviation in the curves for CMB and CMB+CC appear in compared to the curves for CMB+Pantheon and CMB+Pantheon+CC. However, for higher multipoles, we cannot distinguish between the curves.  Finally, we consider the general scenario with $m$ as a free parameter and plot the CMB TT spectra in Fig. \ref{fig-bvf2-cmbmp_general} using the best-fit values of the parameters from all the observational datasets.  We notice that compared to the previous case (BVF2 with $m  =0$), the CMB TT spectra exhibit similar features.

\subsection{Bayesian Evidence}
\label{sec-bayesian}

In the previous subsections we have mainly focused on the observational constraints extracted from different variants of the bulk viscous scenarios. An important question that naturally arises in this context is that, how efficient are the present bulk viscous models compared to the standard $\Lambda$CDM model. The answer can be found by calculating the Bayesian evidence values of the bulk viscous models with respect to this standard $\Lambda$CDM model. The calculation can be done through the publicly available code \texttt{MCEvidence} 
\cite{Heavens:2017hkr,Heavens:2017afc}\footnote{One can freely access this code from the link: 
\href{https://github.com/yabebalFantaye/MCEvidence}{github.com/yabebalFantaye/MCEvidence}.}. Now, one can introduce the so-called Jeffreys scale that quantifies the strength of evidence of the underlying model against the reference model. The Jeffreys scale deals with the values of  $\ln B_{ij}$, where $B_ij$ is the Bayesian evidence of the reference scenario $\Lambda$CDM ($M_i$), with respect to the bulk viscous cosmic model ($M_j$) \cite{Kass:1995loi}.  For $0 \leq \ln B_{ij} < 1$, a Weak evidence; for $1 \leq \ln B_{ij} < 3$, a Definite/Positive evidence; for $3 \leq \ln B_{ij} < 5$, a Strong evidence; and for $\ln B_{ij} \geq 5$, a Very Strong evidence for the $\Lambda$CDM model against the bulk viscous scenario is signaled.  Following \cite{Heavens:2017hkr,Heavens:2017afc} we computed the values of $\ln B_{ij}$ for all such variants of the bulk viscous model with respect to the $\Lambda$CDM model. The values of $\ln B_{ij}$ are summarized in Table \ref{tab:bayesian}  for all possible observational datsets. From Table \ref{tab:bayesian} it is clear that $\Lambda$CDM is favored over the bulk viscous scenarios.  

\begin{table}
\begin{tabular}{ccccccccc}
\hline 
Dataset & Model & $\ln B_{ij}$ & \\
\hline 
CMB & BVF1 ($m =0$)  & $3.4$ \\
CMB+CC & BVF1 ($m =0$)  & $5.7$ \\
CMB+Pantheon & BVF1 ($m =0$) & $2.6$ \\
CMB+Pantheon+CC & BVF1 ($m =0$) & $2.9$ \\
\hline 

CMB & BVF1 ($m$ free)  & $5.1$ \\
CMB+CC & BVF1 ($m$ free)  & $7$ \\
CMB+Pantheon & BVF1 ($m$ free) & $2.9$ \\
CMB+Pantheon+CC & BVF1 ($m$ free) &  $3.6$ \\

\hline\hline 

CMB & BVF2 ($m =0$)  & $4.7$ \\
CMB+CC & BVF2 ($m =0$)  & $7.2$ \\
CMB+Pantheon & BVF2 ($m =0$) & $3.6$ \\
CMB+Pantheon+CC & BVF2 ($m =0$) & $4.3$ \\
\hline 

CMB & BVF2 ($m$ free)  & $6.6$ \\
CMB+CC & BVF2 ($m$ free)  & $9.1$ \\
CMB+Pantheon & BVF2 ($m$ free) & $4.1$ \\
CMB+Pantheon+CC & BVF2 ($m$ free) &  $5.2$ \\
\hline 
\end{tabular}
\caption{The table displays the values of $\ln B_{ij}$ for all the observational datasets and for all variants of the bulk viscous models. Here, $\ln B_{ij} =  \ln B_i - \ln B_j$ ($i$ represents the reference model $\Lambda$CDM and $j$ is for
the underlying model). }
\label{tab:bayesian}
\end{table}

\section{Summary and Conclusions}
\label{sec-conclu}

We have considered a  unified dark fluid endowed with 
bulk viscosity in a spatially flat Friedmann-Lema\^{i}tre-Robertson-Walker (FLRW) universe where the coefficient of the bulk viscosity has a power law evolution: $\eta (\rho_D) = \alpha \rho_d^m$ ($\alpha >0$ and $m$ is a free parameter) and $p_D = (\gamma-1) \rho_D$, $\gamma \in \mathbb{R}$ being the barotropic state parameter. So, one can realize that the above choice for the bulk viscous coefficient automatically includes a number of models, specifically models with fixed $m$. 
For $\gamma =1 $, we rename the scenario BVF1 while for free $\gamma$, we recognize the bulk viscous scenario as BVF2. For $m =0$, one can realize a constant bulk viscous model.  Thus, in order to include the specific cases with $m =0$, 
both the scenarios (i.e., BVF1 and BVF2) have been further classified as (i) the case with $m =0$ [BVF1 ($m=0$), BVF2 ($m=0$)], representing the  constant bulk viscous scenario and (ii) the case for free $m$ [BVF1 ($m$: free), BVF2 ($m$: free)], which is the most general bulk viscous scenario in this work. Thus, essentially we consider four different bulk viscous scenarios and constrain all of them using the observational datasets from CMB, Pantheon sample of Supernovae Type Ia, and the Hubble parameter measurements from the cosmic chronometers. 

For the constant bulk viscous scenarios  BVF1 ($m=0$) and BVF2 ($m=0$), the results of which are summarized in Tables \ref{tab:bvf1-m0} and \ref{tab:results-BVF2-m0} respectively,   
we find that the parameter $\beta $ quantifying the observational evidence of the bulk viscosity is strictly nonzero at  several standard deviations. The model BVF2 ($m=0$) has an additional observational feature that is absent in BVF1 ($m=0$). We find that for the combinations CMB+Pantheon and CMB+Pantheon+CC, the tension on $H_0$ is released within $3\sigma$. In fact, the combination CMB+Pantheon is much effective to reconcile this tension (see Table \ref{tab:results-BVF2-m0}). 

Concerning the general bulk viscous scenarios with free $m$ parameter, the results are summarized in Table \ref{tab:results-BVF1} (BVF1 with free $m$) and Table \ref{tab:results-BVF2} (BVF2 with free $m$).  For BVF1 ($m$ free), we find a strong anticorrelation between the free parameter $m$ and the quantifying bulk viscous parameter $\beta$. Precisely, we find that $\beta$ is nonzero at more than $2$ standard deviations for all the datasets. Regarding the $m$ parameter, although for CMB alone $m =0$ is allowed within 68\% CL, but for other datasets, $m \neq 0$ becomes strongest. In addition, we see that for the addition of Pantheon to CMB reduces the 
tension on $H_0$ weakly. For the BVF2 ($m$ free) scenario, we have similar observations as in BVF1 ($m$ free). We find that although for CMB alone data, $m =0$ is certainly the case, but for other datasets, $m \neq 0$ is strongly preferred. The parameter $\beta$ quantifying the bulk viscosity in the universe, we find its positive evidence for all the observational datasets at more than $2$ standard deviations. Additionally, for the CMB+Pantheon and CMB+Pantheon+CC datasets, this specific bulk viscous scenario has the ability to reduce the $H_0$ tension weakly. Thus, considering the observational limits on both $\beta$ and $m$ for both BVF1 and BVF2, a strong indication of a non-zero bulk viscous scenario (since $\beta \neq 0$) apart from the constant bulk viscous coefficient (since $m \neq 0$ except for CMB alone), 
is supported. We also perform a Bayesian evidence analyses for all the bulk viscous scenarios shown in Table \ref{tab:bayesian} aiming to compared them with respect to the standard $\Lambda$CDM model.  Our analyses report that $\Lambda$CDM is favored over the bulk viscous scenarios, at least for the employed observational datasets.

To conclude, perhaps along with other findings, the best finding is to explore the  ability of the bulk viscous scenarios to reduce the $H_0$ tension weakly. We use weakly in the sense that the tension is released under $3\sigma$ CL. 
In summary, we see that the bulk viscous scenarios might be able to compete with other cosmological models in which an additional constraint in terms of either the inclusion of phantom dark energy equation of state \cite{DiValentino:2016hlg,DiValentino:2017zyq,DiValentino:2017rcr,Vagnozzi:2018jhn,Yang:2018qmz} or the nonzero  interaction \cite{Kumar:2017dnp,DiValentino:2017iww,Yang:2018euj,Yang:2018uae,Pan:2019jqh}, are necessary to release the $H_0$ tension.

\section{Acknowledgments}
The authors thank the referee for some essential comments that helped us to improve the manuscript. 
The authors also thank Prof. J. D. Barrow for some very useful discussions while working on this article. 
WY has been supported by the National
Natural Science Foundation of China under Grants No.  11705079 and No.  11647153. SP has been supported by the Mathematical Research Impact-Centric Support Scheme (MATRICS), File No. MTR/2018/000940, given by the Science and Engineering Research Board (SERB), Govt. of India.  SP also acknowledges partial support from the Faculty Research and Development Fund (FRPDF) Scheme  of Presidency University, Kolkata, India.  E.D.V. acknowledges the support from the European Research Council in the form of a Consolidator Grant with number 681431.

\end{document}